\documentclass[10pt]{iopart}
\usepackage{graphicx}
\usepackage{latexsym}
\usepackage{dcolumn}
\usepackage{bm}
\usepackage{color}
\usepackage{amssymb}
\usepackage[normalem]{ulem}
\usepackage{mathtools}
\usepackage[utf8]{inputenc} 
\usepackage[T1]{fontenc}

\newcommand{\be}{\begin{equation}}
\newcommand{\ee}{\end{equation}}
\newcommand{\bea}{\begin{eqnarray}}
\newcommand{\eea}{\end{eqnarray}}
\newcommand{\bse}{\begin{subequations}}
\newcommand{\ese}{\end{subequations}}

\newcommand{\erf}{\mbox{erf}}

\renewcommand{\e}{\epsilon}

\renewcommand{\comment}[1]{}

\begin{document}

\paper[]{Infinite-temperature thermostats by energy localization in a nonequilibrium setup}
\date{\today}

\author{Michele Giusfredi$^{1,2,3}$, Stefano Iubini$^{2,3}$, Antonio Politi$^{2,4}$, Paolo Politi$^{2,3}$}

\address{$^{1}$ Dipartimento di Fisica e Astronomia, Universit\`a di Firenze,
via G. Sansone 1 I-50019, Sesto Fiorentino, Italy}
\address{$^{2}$ Istituto dei Sistemi Complessi, Consiglio Nazionale
delle Ricerche, via Madonna del Piano 10, I-50019 Sesto Fiorentino, Italy}
\address{$^{3}$ Istituto Nazionale di Fisica Nucleare, Sezione di Firenze, via G. Sansone 1 I-50019, Sesto Fiorentino, Italy}
\address{$^{4}$ Institute for Complex Systems and Mathematical Biology
           University of Aberdeen, Aberdeen AB24 3UE, United Kingdom}
\ead{michele.giusfredi@unifi.it,stefano.iubini@cnr.it,\\a.politi@abdn.ac.uk,paolo.politi@cnr.it}

\begin{abstract}
Some lattice models having two conservation laws may display an
equilibrium phase transition from a homogeneous (positive temperature - PT) to a
condensed (negative temperature) phase, where a finite fraction of the energy is localized in
a few sites. 
We study one such stochastic model in
an out-of-equilibrium setup, where the ends of the lattice chain are
attached to two PT baths.
We show that localized peaks may spontaneously emerge, acting as
infinite-temperature heat baths.
The number $N_b$ of peaks is expected to grow in time $t$ as
$N_b \sim \sqrt{\ln t}$, as a consequence of an effective freezing of the dynamics. 
Asymptotically, the chain spontaneously subdivides into three intervals:
the two external ones lying inside the PT region;
the middle one characterized by peaks superposed to a background lying
along the infinite-temperature line.
In the thermodynamic limit, the Onsager formalism allows determining
the shape of the whole profile.







\end{abstract}


\submitto{Journal of Statistical Mechanics: theory and experiment}

\maketitle

\section{Introduction}

Simple non-trivial models have always played an important role in statistical mechanics, 
because they allow to highlight special features and phenomena focusing on the minimal 
ingredients that are necessary and sufficient to generate a given behaviour: 
the Ising~\cite{Ising_McCoy} and the driven lattice gas models~\cite{Schmittmann_Zia} are 
the most famous examples in the context of equilibrium and out-of equilibrium phase transitions, respectively.
Another example, less known but more relevant in the present context, is the model introduced by 
Kipnis, Marchioro, and Presutti (KMP) in 1982~\cite{kmp82}, which allowed to prove the validity 
of the diffusion equation. In that model, a positive quantity $c_i$ (mass) is 
associated to each site $i$ of a one dimensional lattice, and dynamics proceeds by exchanging mass between 
neighbouring sites $(i,i+1)$ in such a way that the sum $(c_i + c_{i+1})$ is kept constant. 
More recently, a model has appeared with a much richer phenomenology, but conceptually similar to the KPM model.
In this case~\cite{Iubini2013_NJP}  there are two diffusing conserved quantities ($c_i,\e_i$),  
one of which (the energy $\e_i$, let us say) is locally the square of the other, $\e_i = c_i^2$.
This model displays an equilibrum phase transition between a homogeneous phase and a condensed/localized
phase, where a single site hosts a finite fraction of the whole, macroscopic energy.
For this reason, the model was called C2C~\cite{GIP21}: 
condensation with (because of) two conservation laws.
A crucial feature of the model is that both mass and energy variables are bounded from below. 
Had we allowed $c_i$ to take negative values, as, e.g., done in Ref.~\cite{Miron2019}, where $c_i$ plays
the role of linear momentum, we would have
devised yet another stochastic model of heat conductivity.

The C2C model can be seen as an extension of KPM to the case of coupled transport, but it is also related
to different classes of models: the Discrete NonLinear Schr\"odinger (DNLS) equations~\cite{kevrekidis09,JSP_DNLS} 
and the Zero-Range Processes (ZRP)~\cite{Evans2005_JPA}.
In fact, the DNLS equation reduces to C2C in proximity of the critical line of the condensation 
transition~\cite{Iubini2013_NJP,Gradenigo2021_JSTAT,arezzo22} and one can identify tall breathers
arising in the DNLS equation with the condensation peaks discussed in this paper.
Moreover, the C2C corresponds to a ZRP-like process~\cite{Gradenigo2021_JSTAT,GIP21},
where energy is the only conserved quantity and mass conservation is replaced by a Weibull distribution of
energy on each site, $f(\e) = (\lambda/2\sqrt{\e})e^{-\lambda\sqrt{\e}}$.

Here, we assume and consider only a quadratic relationship between $\e_i$ and $c_i$,%
but this is not a crucial constraint:
any convex function $\e_i = F(c_i)$ would give the same phenomenology~\cite{GIP24}. 
The equilibrium properties of the C2C model are well known and can be easily summarized in terms of the 
mass density $a=(1/N)\sum_i c_i$ and the energy density $h=(1/N) \sum_i c_i^2$,
where $N$ is the total number of lattice sites.
The homogeneous phase, $a^2\le h \le 2a^2$, can be equally described within the microcanonical 
and the grand canonical ensembles; the two curves $h=a^2$ and $h=2a^2$ represent the states at
zero temperature and infinite temperature, respectively.
The region $h>2a^2$ identifies the localized phase and its infinite-time behaviour can be described only
within the microcanonical ensemble. However, close enough to the critical line, the homogeneous phase
is metastable on long time scales and an effective grand canonical description is possible~\cite{IP24}.
In both ensembles, absolute negative temperatures (NT)~\cite{Baldovin_2021_review} appear in the critical region of the
localized phase.

Recently~\cite{gotti22}, it has been found that the localization process may appear in an out-of-equilibrium setup
where the ends of the lattice chain are attached to two thermal baths (reservoirs) operating
at strictly positive temperatures (PT).
In fact, although each bath thermalizes (in its vicinity) the system into a homogeneous phase,
the Joule effect induced by the coupled transport of mass and energy leads to such
a large temperature increase, that it becomes negative!
Since coupled transport is crucial for non-equilibrium condensation to appear, 
it is obvious that ZRP-like models cannot exhibit this phenomenon~\cite{Levine2005}. 

When passing to an out-of-equilibrium setup, tiny details of the dynamics may start playing
a crucial role, giving rise to possibly different scenarios.
One of them is the mobility of the energy peaks appearing in the localized phase.
They are energy storages which may or may not travel. 
If they diffuse, energy can be eventually dissipated in correspondence of the heat baths, leading to
a nonequilibrium steady state (NESS) with peaks that are continuously created and destroyed~\cite{GIP24}.

If, instead, energy peaks are pinned (spatially blocked), nothing prevents their energy load from increasing forever:
this is the scenario we are going to discuss in detail in this paper.
In particular, we will show that the coexistence of point singularities and low-energy regions
formally violates the NESS constraints
and gives rise to an unusual transport regime.

Peaks of arbitrarily large energy may appear and be sustained also at equilibrium, in a microcanonical setup where
energy and mass are  conserved~\cite{Rumpf2004,Gradenigo2021_EPJE}. 
In fact, it is known that in the condensed phase, a single peak either absorbs or emits energy until
the background is in an infinite-temperature equilibrium state (this is the way to maximize the entropy).
Hence, one can conclude that energy peaks act as infinite-temperature reservoirs.

\begin{figure}[!ht]
        \centering
        \includegraphics[width=0.9\columnwidth,clip=yes]{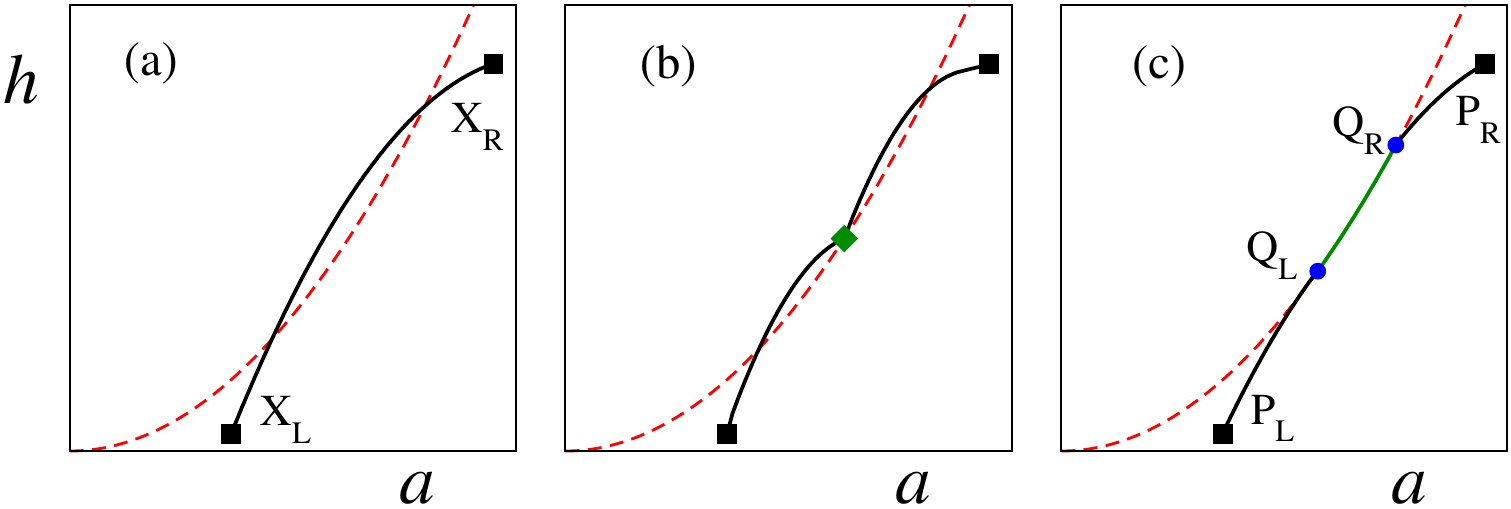} 
        \caption{
		Sketch of the parametric curve (full line)  for an out-of-equilibrium setup when
		heat baths $(X_L,X_R)$ are in the homogeneous phase, i.e., below the critical line
		(red, dashed line). 
	    (a)~The metastable regime, when no peak has yet appeared;
	    (b)~The emergence of an immobile peak (green diamond) bends the parametric profile
		towards the critical line;
		(c)~The asymptotic configuration: The positive-temperature sections of the
		parametric profile end tangentially to the critical line (blue dots) and 
		the inner part from $Q_L$ to $Q_R$ is squashed on top of it.
        }
        \label{fig:sketch}
    \end{figure}

Within the out-of-equilibrium setup, the evolution of a system entering the condensed phase is sketched
in Fig.~\ref{fig:sketch}, where the state of the $i$th lattice site is represented as a point
of coordinates $( \langle c_i\rangle,\langle c_i^2\rangle)$
in the plane $(a,h)$, where
the angular brackets denote a finite-time average.
As a result, the current state of the system can be viewed as a parametric curve connecting the two reservoirs.
At short times (see left panel), the system settles in a metastable state, whose curve crosses
the critical line, suggesting that the conditions for the occurrence of an  out-of-equilibrium condensation are met.
At later times, a first peak emerges (see central panel), and the related curve bends towards the critical line,
indicating that the peak acts as an infinite$-T$ reservoir.
As we will see, a single peak is not sufficient to absorb the extra-energy pumped in the system. In fact,
large part of this paper is devoted to understanding how additional peaks form, and how the parametric
curve progressively changes.
In the right panel of Fig.~\ref{fig:sketch} we depict the expected asymptotic form,
once a large number of peaks has appeared.

The paper is organized as follows.
The model and the main quantities used throughout the article are defined in Sec.~\ref{sec.model}.
In the following Sec.~\ref{sec.phenomenology}, we describe the effects of high-energy peaks
on the macroscopic nonequilibrium state
(either when peaks rise spontaneously or when they are externally imposed on a given site). 
Since the complexity of the evolution stems from the dynamics of the segment of parametric curve invading the condensed phase,
we devote Sec.~\ref{sec.metastable}, \ref{sec.long-time} to a quantitative analysis of a setup where
both thermostats act on the critical line (we shall speak of critical reservoirs).
In particular, Sec.~\ref{sec.metastable} is devoted to the analysis of the pseudo-statonary metastable regime 
(i.e. before the emergence of new peaks), while in Sec.~\ref{sec.long-time}, we discuss how new peaks progressively emerge.
In Sec.~\ref{sec.PNpaths}, we return to the general case, and show how the asymptotic path can be constructed,
when both thermostats are located in the positive temperature region.
Finally, in Sec.~\ref{sec.discussion} we provide a critical summary of our results  and discuss  possible extensions to other models.

\section{The model}
\label{sec.model}

The C2C model is defined in terms of two conservation laws. The mass and energy densities $(a,h)$ define the
microcanonical equilibrium state of an isolated system. If mass and energy exchanges with an external reservoir are allowed, the system can reach an equilibrium state characterized by a temperature $T$ and a chemical potential $\mu$. In the homogeneous phase, the grand canonical description is well defined with a grand canonical partition function,
\be
Z(\beta, \mu) = \left( \int_0^{+\infty} d c \, e^{-\beta(c^2 - \mu c)} \right)^N \equiv z(\beta, \mu)^N,
\ee 
where $\beta = 1/T$ is the inverse temperature. 
The local mass is therefore distributed according to 
$\rho(c) = 1/z(\beta, \mu) e^{-\beta(c^2 - \mu c)}$.
The  microcanonical $(a,h)$ and the grand canonical $(\beta,\mu)$  parameters can be 
mapped on to one another using the relations~\cite{Szavits2014_PRL,Szavits2014_JPA,gotti22}
\be
a =  \frac{\mu}{2} + \frac{1}{\sqrt{\pi \beta}} \frac{e^{-\beta \mu^2/4}}{1 + \erf \left( \frac{\sqrt{\beta} \mu}{2}\right)},
\label{mapping_a_mub}
\ee
\be
h =  \frac{1}{2 \beta} + \frac{1}{2} a \mu.
\label{mapping_h_mub}
\ee

The critical line, $h_c = 2 a^2$, is characterized by a vanishing $\beta$ and a diverging
chemical potential $\mu$, in such a way that the product $m = \beta \mu$ remains 
a finite negative constant, $m \to -1/a$. On the critical line  the local mass distribution reduces to the exponential distribution, $\rho(c) = 1/a \, e^{-c/a}$.

While the equilibrium phase transition of the  C2C model depends only on the value of the 
two conserved quantities, the relaxation towards equilibrium  and out-of-equilibrium properties 
depend also on the topology of the system and on the microscopic dynamical rule defining time evolution.
The out-of-equilibrium setup we consider here is a one-dimensional lattice of size $N$, whose outermost sites are connected with two different heat baths, which impose the thermodynamic parameters $X_L = (\beta_L, m_L)$ and  $X_R = (\beta_R, m_R)$ on the left and right ends of the chain, respectively. When $X_L = X_R$, the system can relax towards an equilibrium state, while for $X_L \neq X_R$ the chain is forced into a non-equilibrium state in which mass and energy are transported from one reservoir to the other.

The system is evolved according to a stochastic algorithm~\cite{JSP_DNLS}: at each step we randomly extract an 
integer $k \in [2,N-1]$. If $k \in [3,N-2]$, we update the triplet of adjacent sites $(k-1, k, k+1)$ and substitute their local masses  $\vec{I} = (c_{k-1},c_{k},c_{k+1})$ with new ones $ \vec{F} = (c_{k-1}',c_{k}',c_{k+1}')$ so that both their sum and the sum of their squares do not change. The mass triplets can be interpreted as points in a tridimensional space and the two constraints define a circumference, parametrized by an angle. The new triplet $\vec{F}$ is then chosen among the admissible configurations by extracting an angle from a uniform distribution. 
Under this assumption, it is possible to prove that detailed balance is satisfied (see
\ref{app.triplet_rule} for more details).

Since the local masses are non-negative, the allowed configurations may reduce to three disjoint arcs, 
rather than lying on the entire circle. 
In Ref.~\cite{GIP24}, the new point $\vec F$ was chosen in any arc, thus allowing the peak to diffuse.
Here, as well as in \cite{iubini23}, we select the triplet $\vec{F}$ within the same arc containing $\vec{I}$. 
This constraint implies that when one of the masses is sufficiently larger than the others, 
the peak does not change its position within the triplet. 
The resulting dynamics is akin to that of the DNLS model, where breathers indeed do not diffuse.

The interaction with the heat baths is treated as follows.
When $k=2$ ($k=N-1$) is randomly extracted, a new $c_1$ ($c_N$) value is generated
(before updating the triplet)
by extracting a value from the equilibrium mass distribution $\rho_{L}(c)$ ($\rho_{R}(c)$),
defined according to the thermodynamic parameters $X_{L}$ ($X_R$) of the heat bath~\cite{iubini23, GIP24}.
The equilibrium distribution is well defined only for strictly positive temperatures (including the
critical case $T=\infty$).
It is possible to define an effective equilibrium distribution also for NTs slightly above
the critical line (see, Ref.~\cite{IP24}), where the corresponding regimes are
metastable. In this paper we consider only PT thermostats,
since, this way, we avoid the additional complication of boundary induced instabilities.
A general discussion of NT thermostats can be found in~\cite{Baldovin_2021_review}.

\comment{
The interaction with the heat baths is treated as follows.
When we extract $k=2$ ($k=N-1$), before updating the triplet, a new $c_1$ ($c_N$) value is generated
by extracting a value from the equilibrium mass distribution $\rho_{L}(c)$ ($\rho_{R}(c)$),
defined according to the thermodynamic parameters $X_{L}$ ($X_R$) of the heat bath~\cite{iubini23, GIP24}.
Finally, time is measured in Monte Carlo time units, which correspond to $N$ local updates.
}

Fluxes are the most important observables for the characterization of NESSs and metastable states.
A local definition of flux can be given by referring to the amount of either mass or energy travelling
across a given link.
With reference to the pair of adjacent sites $(k,k+1)$, a variation $J^{k,k+1}_a$ of mass
($J^{k,k+1}_h$ of energy) can occur only when two specific triplets are selected, namely
$T_k = (k-1,k,k+1)$  or $T_{k+1} = (k,k+1,k+2)$. Depending on which triplet is selected, the instantaneous
mass flux can be defined as
\begin{equation}
 J^{k,k+1}_a(T_k) = c'_{k+1}-c_{k+1},  \qquad   \quad
 J^{k,k+1}_a(T_{k+1}) = c_{k}-c'_{k}      
\end{equation}
where we are adopting the convention that a positive value corresponds to a flux from left to right.
Analogously, the energy flux is defined as
\begin{equation}
 J^{k,k+1}_h(T_k) = \epsilon'_{k+1}-\epsilon_{k+1},  \qquad   \quad
 J^{k,k+1}_h(T_{k+1}) = \epsilon_{k}-\epsilon'_{k}      
\end{equation}
For $k=1$, only triplet $T_2$ can occur and $J^{1,2}_a = c_1-c'_1$, while  $J^{1,2}_h = \epsilon_1-\epsilon'_1$.
For $k=N-1$, only triplet $T_{N-1}$ can occur and $J^{N-1,N}_a = c'_N-c_N$, while $J^{N-1,N}_h = \epsilon'_N-\epsilon_N$.

Finally, the average mass (energy) flux $J_a$ ($J_h$) is determined by dividing the amount of mass (energy)
flowed over a given time $\tau$ by the amount itself of elapsed time.%
\footnote{Time is measured in Monte Carlo time units, 
which correspond to $N$ local updates.}

It is also useful to define the mass- and energy- current asymmetries,  
\be
\label{eq:as_ener_curr}
  \Delta J_{a,h} = J_{a,h}^{(L)} - J_{a,h}^{(R)} \; 
\ee
where $J_{a,h}^{(L)} = \langle J^{1,2}_{a,h}\rangle$ and
$J_{a,h}^{(R)} = \langle J^{N-1,N}_{a,h}\rangle$.

If the system evolves towards a NESS, both mass and energy flux asymmetries vanish, 
$\Delta J_{a,h} \to 0 $,
and asymptotically $J_a^{(L)} = J_a^{(R)}$ and  $J_h^{(L)} = J_h^{(R)}$. 
A NESS is always reached when the dynamics allows peaks to diffuse~\cite{GIP24}, both when the parametric profile $h_i(a_i)$ remains below the critical line and when it enters the NT region. As we will show in the next sections, this is no longer true when the energy peaks are pinned.

Finally, we introduce the formalism of linear response theory~\cite{lepri2016}, which is applicable for large system sizes $N$, under the assumption of a local equilibrium, with small currents and gradients of thermodynamic quantities.
In this limit it is useful to consider rescaled fluxes $j_{a,h} = N J_{a,h}$, which asymptotically
do not depend on $N$.
Furthermore,  the spatial index $i$ can be treated as a continuous variable, introducing $x=i/N$, which corresponds
to a unit length for the whole chain.

Local mass and energy currents are  
related to the thermodynamic forces, $\partial_x \beta$ and $\partial_x m$, via the
linear coupled-transport equations~\cite{livi17,iubini23}
\begin{equation}
    j_a = - L_{aa}(\beta, m) \partial_x m + L_{ah}(\beta, m) \partial_x \beta 
    \label{LCTeq1}
\end{equation}

\begin{equation}
    j_h = - L_{ha}(\beta, m) \partial_x m + L_{hh}(\beta, m) \partial_x \beta 
    \label{LCTeq2}
\end{equation}
where $L_{r s}$, with $r,s = \{a,h\}$, are the Onsager coefficients.
As shown in~\cite{iubini23}, the linear response approach can be applied in the 
PT region 
for arbitrarily high temperatures and the Onsager coefficients remain well defined and finite even on the critical line. 

The Onsager coefficients are functions of the thermodynamical parameters $(\beta, m)$, but their dependence can
be simplified by exploiting the scaling properties of the model, as shown in Ref.~\cite{iubini23}. 
Defining the auxiliary variable $w = \sqrt{\beta}/m$, we obtain the scaled forms
\be
L_{aa} = |m|^{-2} \overline{L}_{aa}(w), \quad 
L_{ah} = L_{ha} = |m|^{-3} \overline{L}_{ah}(w), \quad 
L_{hh} = |m|^{-4} \overline{L}_{hh}(w), 
\label{eq_OM}
\ee
which imply that on the critical line, where $w =  0$, the Onsager coefficients are simply 
powers of $|m|$, $L_{r s} (\beta = 0,m ) = |m|^{\gamma_{rs}} \overline{L}_{r s} (0)$,
with $\gamma_{aa}=-2$, $\gamma_{ah}=\gamma_{ha}=-3$, and $\gamma_{hh}=-4$.

  \section{Phenomenology}
  \label{sec.phenomenology}
  
In this section we illustrate the effect of a large peak in two distinct setups, 
where both thermostats 
operate inside the PT region. In the former, peaks arise spontaneously because the parametric path enters the NT region; 
  in the latter, a peak  is artificially added on a lattice-site belonging to the PT region to monitor its effect.
    
\begin{figure}[!h]
        \centering
        \includegraphics[width=0.7\columnwidth]{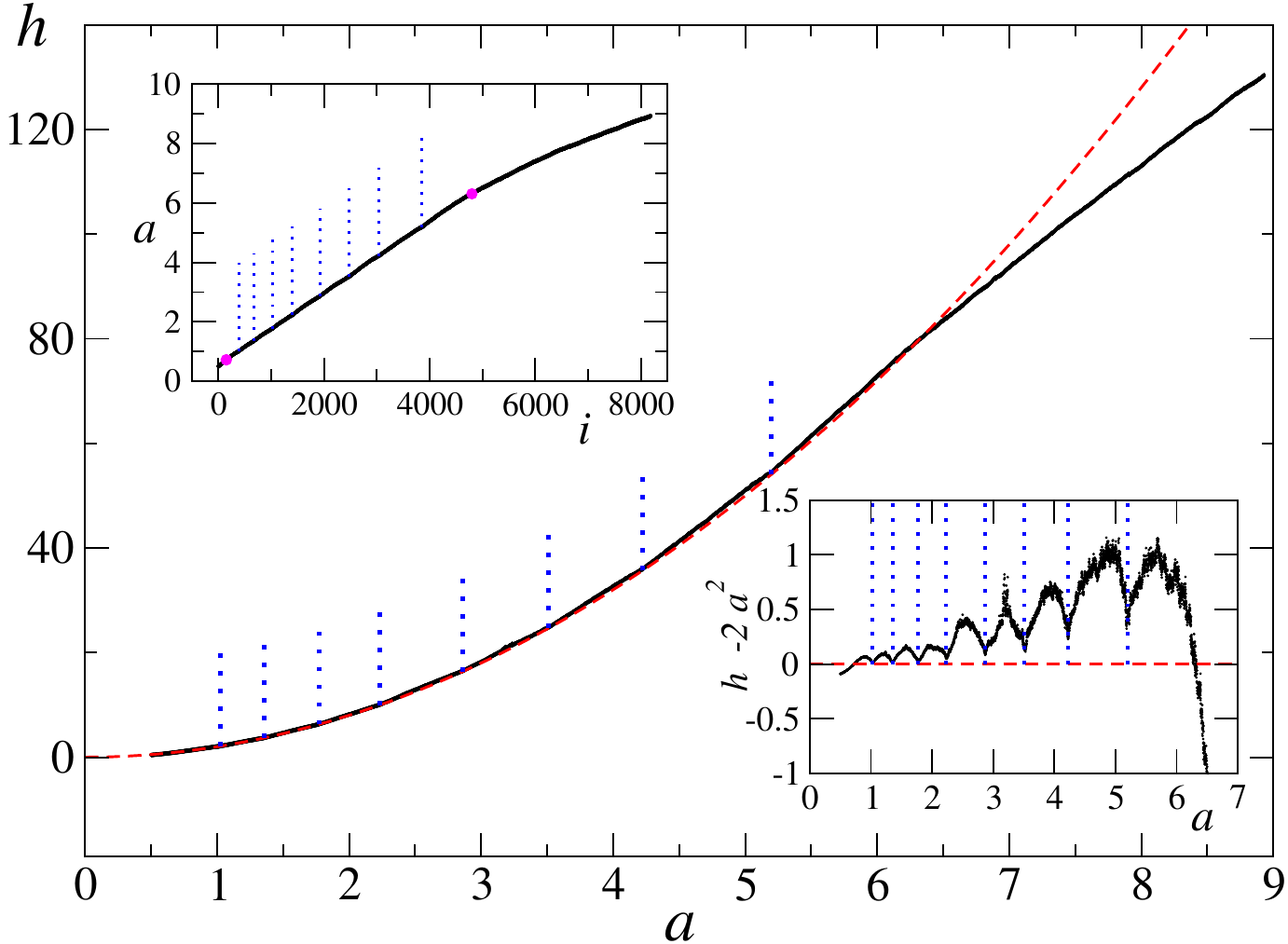} 
        \caption{Parametric profile in the plane $(a,h)$ for a chain of size $N= 8192$ connected with positive temperature heat baths with thermodynamic parameters $(\beta_L, m_L) = (0.884, -0.550)$ and $(\beta_R, m_R) = (0.00286, -0.0286)$. The dashed red line represents the critical curve $h= 2a^2$. The dotted vertical lines indicate the position of the 8 peaks that formed spontaneously. The values of $a$ and $h$ were obtained from an average over $2.3\times 10^8$ time units after the formation of the last peak.
        Bottom right inset: The same parametric profile now plotted in the plane $(a, h - 2a^2)$. 
        Top left inset: The spatial mass profile $a_i$. The magenta dots are the positions of the sites where the parametric profile crosses the critical line. The meaning of the lines in both insets is the same as for the main figure.   
        }
        \label{fig:ah_profile}
    \end{figure}

In Figs.~\ref{fig:sketch}(a,b) we sketched a parametric profile of the first type, 
anticipating the effect of the appearance of a single peak. 
In Fig.~\ref{fig:ah_profile} we plot the ``real'' parametric profile emerging after a long but finite time. 
There, we see that eight peaks (whose position is signaled by the vertical dotted lines) 
have formed.  The overall pattern is better appreciated in the lower inset, where we recognize that also the left reservoir operates at a positive temperature, and that the profile entering the NT region spontaneously segmented into nine pieces, each having a dome shape, except for the 4th one from the right, 
where a new peak was presumably in the process of being generated.
The main message of these simulations is that
the formation of a peak tends to locally anchor the profile to the critical curve, suggesting that peaks act 
as effective infinite-temperature heat baths.
  
\begin{figure}[!h]
        \centering
        \includegraphics[width=0.7\columnwidth]{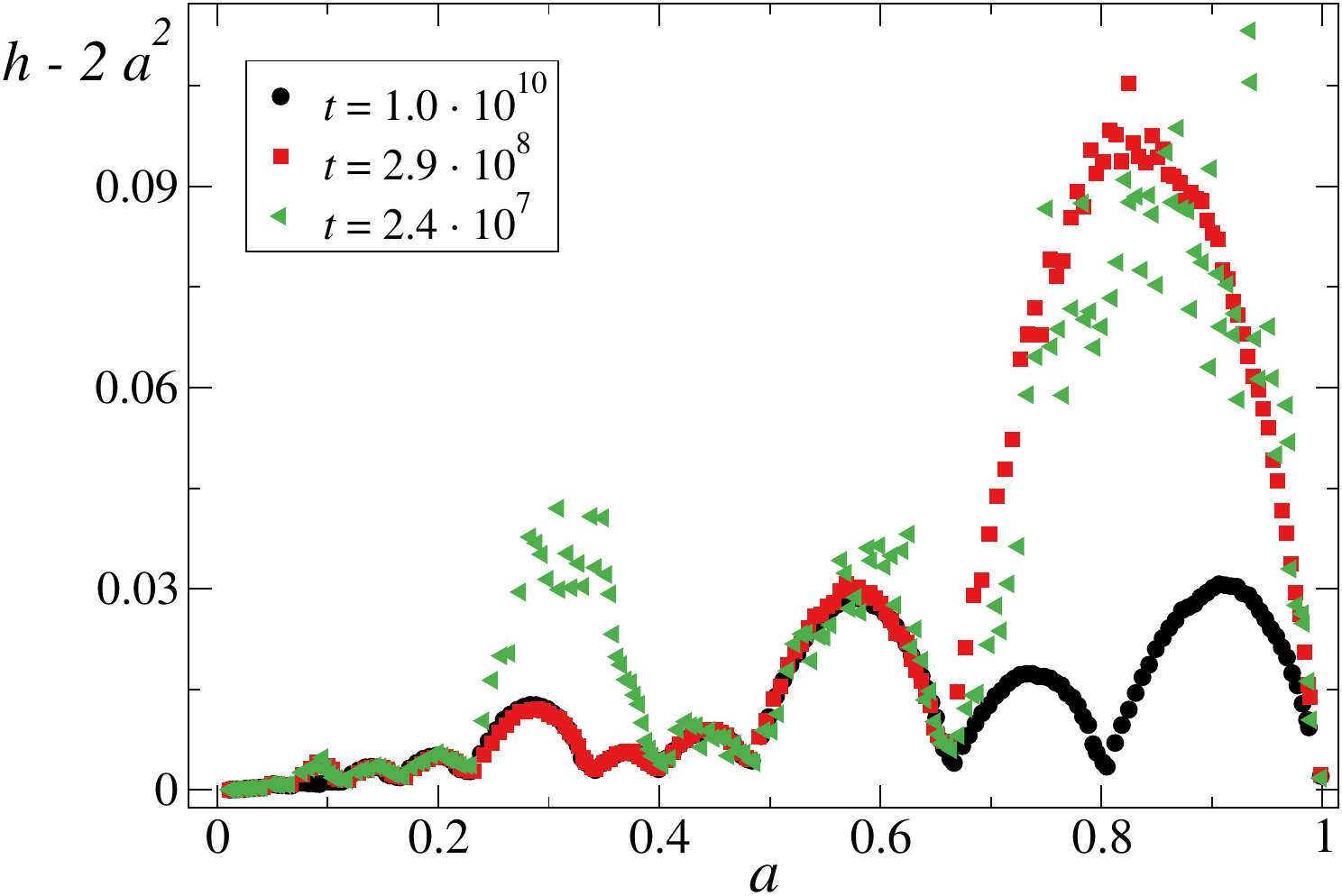} 
	 \caption{Parametric profiles in the plane $(a, h-2a^2)$ for a chain of size $N= 200$ connected with critical heat baths at $a_L = 0.01$ and $a_R = 1$, at different times of the same simulation. The values of $a$ and $h$ are obtained from temporal averages between the appearance of a peak and the next. 
        }
        \label{fig:aDh_profile}
    \end{figure}

The idea is strenghtened in Fig.~\ref{fig:aDh_profile}, where profiles taken at different times are plotted for a setup where
both thermostats operate along the critical line. Peaks (not reported explicitly) are located in correspondence of the 
downward cusps, similarly to the lower inset of Fig.~\ref{fig:ah_profile}. 
Their number increases in time, while the height of the ``domes'' connecting them decreases,
indicating that the profile is closer to the critical line.

Energy peaks also have an effect on spatial profiles, as shown in the upper inset of 
Fig.~\ref{fig:ah_profile} where we plot the mass spatial profile $a_i$.
In the portion of the chain where the parametric profile enters the NT region
such profile deviates very little from a linear profile. 
As the number of peaks increases, the profile become more and more linear, being
exactly linear when the parametric profile lies on the critical curve.
This can be proved analitically in both a discrete and a continuous model, see \ref{app.mep}
and \ref{app.cp}.

\begin{figure}[!h]
\begin{center}
\includegraphics[width=0.7\textwidth,clip]{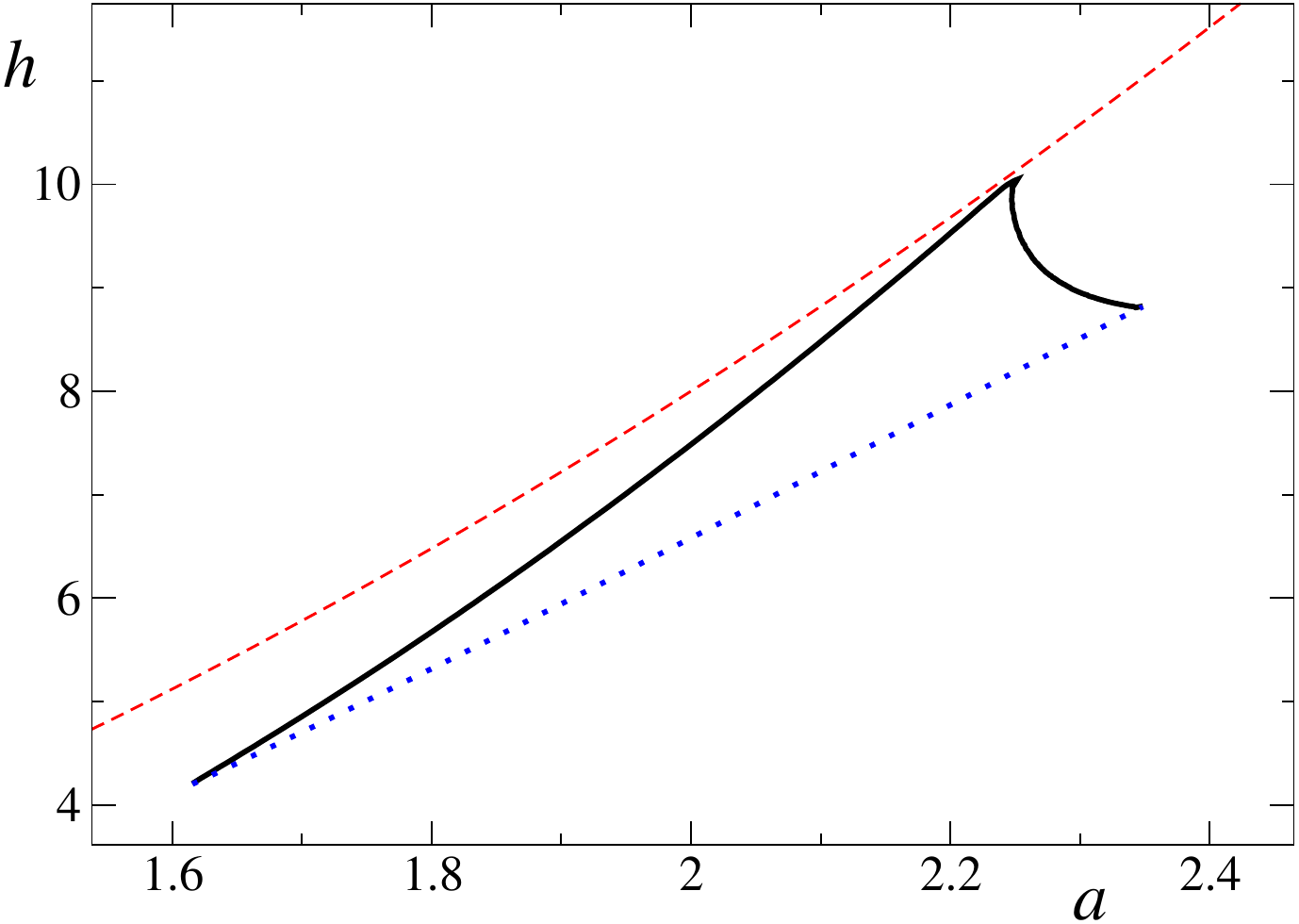}
\end{center}
\caption{
Parametric profiles in the plane $(a, h)$ for a chain of size $N= 200$ connected with positive temperature heat baths with $(\beta_L, \mu_L) = (0.1, -1)$ and $(\beta_R, \mu_R) = (0.05, -1)$. 
The blue dots form the spontaneous parametric profile, while the black full line is the profile 
in case we impose an infinite peak in the center of the chain (at site $i = 100$). The red dashed line is the critical curve $h_c = 2a^2$. The values of $a, h$ have been obtained with temporal averages over a time $t = 2.5 \times 10^9$.
}
\label{fig:1breatherTfin}
\end{figure}

We now discuss a different setup where the parametric path does not enter the NT region: see the 
dotted blue curve in Fig.~\ref{fig:1breatherTfin}.
If we artificially add a peak in the middle of the chain (by formally attributing a very large 
mass to a given site), 
the energy obviously decreases in time as a manifestation of a relaxation to (local) equilibrium. 
However, if the peak mass is infinitely large,\footnote{This hypothesis is a typical artifice,
when dealing with heat reservoirs,
which are supposed to absorb or release arbitrary amounts of energy without
changing their properties.} a new NESS emerges,
whose parametric profile (black full curve) touches the critical line in
correspondence of the peak.

The equivalence between peaks and infinite$-T$ reservoirs can be intuited already from 
the analysis of the microcanonical \textit{equilibrium} state, when $h>h_c(a)$: 
in the thermodynamic limit, a single peak collects the energy $E_b = N(h - h_c(a))$, while the rest of the system is kept at infinite $T$~\cite{Rumpf2004}:
a picture that is perfectly consistent with the idea that a peak acts as an infinite$-T$ heat bath.
But why does this happen? In a first step towards the comprehension of the phenomenon, we consider the evolution of a triplet, 
where the mass $b$ of one of the three sites is much greater than the other two masses, $c_1$ and $c_2$. 
In this limit, $b\gg c_{1,2}\simeq \mathcal{O}(1)$, the mass variation of the peak is very small,
$\delta b \simeq 1/b$ (see Ref.~\cite{JSP_DNLS} and \ref{app.triplet_b}),
so that the variation of its energy, $\delta(b^2)$, is of order one, 
allowing to compensate the energy variation of the other two sites. 
The key point is that a negligible mass variation of the peak implies that the evolution of $c_1$ and $c_2$ is mostly determined by the mass conservation law, $\delta c_1 + \delta c_2 = 0$, while energy conservation determines $\delta b$ through the relation $\delta (b^2) + \delta (c_1^2) + \delta (c_2^2)=0$, where $\delta (x^2) \equiv (x+\delta x)^2 - x^2$.\footnote{In the setup of Fig.~\ref{fig:1breatherTfin}, where the whole system would stay at $PT$, 
the imposed peak would have a negative average value $\langle\delta b\rangle$, signaling that in such conditions no peak can be naturally sustained. This is not problematic, as explained in the
previous note.}

Therefore, introducing a peak into the system means that neighbouring sites evolve according to the sole law of mass conservation: this implies an exponential distribution of the mass, which is the typical feature of infinite$-T$ states. 
The same conclusion can be rationalized  both
from a grand canonical and a microcanonical point of view. From a grand canonical point of view, the statistical weight $\exp(-\beta (c^2 -\mu c))$ reduces to $\exp(\beta\mu c)$, corresponding to $\beta=0$ (and $\beta\mu$ finite). From a microcanonical point of view, the exponential distribution implies that the second moment of $a_i$ is twice the square of the first moment, $h=2a^2$, therefore placing the site on the critical curve. 

\begin{figure}[!h]
	 \begin{center}
                \includegraphics[width=0.7\columnwidth]{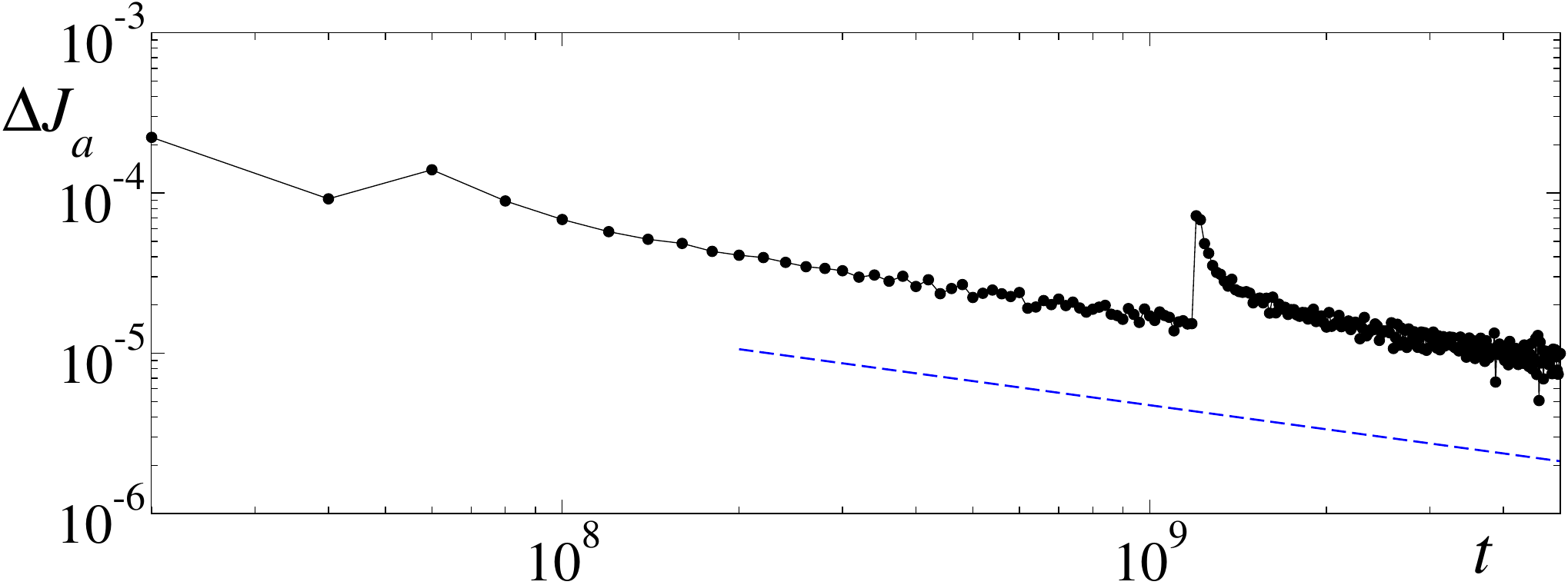} 
	 \end{center}
        \caption{
		Asymmetry of the mass flux $\Delta J_a$  \textit{vs} time for a system of size $N=32$ connected to critical heat baths at $a_L= 1$, $a_R=5$.  The flux asymmetry at time $t$ is computed by averaging over the 
time interval $[t-\tau, t]$, with $\tau = 2 \times \, 10^7$. The sudden jump of the asymmetry $\Delta J_a$ at time $t = 1.2 \times 10^9$ is due to the appearance of a new peak on the system. The blue dashed line is the reference $1/\sqrt{t}$.
        }\label{fig:flusso_massa_n32}
    \end{figure}

It is now interesting to monitor the time-averaged mass and energy currents, respectively $J_a$ and $J_h$. If the system attains a steady state, it is characterized by constant currents in time and space: therefore, the currents $J_{a,h}$ exchanged with the left reservoir are equal to the currents $J_{a,h}$ exchanged with the right reservoir. This is exactly what happens if peaks are allowed to diffuse,
in which case the currents have been determined analytically~\cite{JSP_DNLS}: $J_a=-2(a_R - a_L)/N$ and $J_h = -2(h_R - h_L)/N$. If peaks are instead  pinned, their unlimited growth hinders the trasmission of energy between the heat baths and creates an asymmetry between inflows and outflows. However, since the mass variation of a peak is asymptotically negligible, we expect an  asymmetry to occur
only for the energy current. This is confirmed in Fig.~\ref{fig:flusso_massa_n32}, where the asymmetry $\Delta J_a = J_a^{(L)} - J_a^{(R)}$,
plotted as a function of time, vanishes upon increasing time. 
At long times, $\Delta J_a$ decreases as $1/\sqrt{t}$. This can be related to the results found with the analytical treatment of the triplet: in this case the mass-flux asymmetry is equal to the average mass variation of the peak, 
that decreases as $1/b$ with $b$~\cite{JSP_DNLS}. Therefore, the average growth rate
of the peak energy $b^2$ becomes constant when the  mass $b$ is large, 
$b^2(t)\sim t$, which implies $b\sim \sqrt{t}$, so that  $\Delta J_a \sim 1/\sqrt{t}$. In a large system with multiple growing peaks, $\Delta J_a$ is given by the sum of the average mass 
variations of all the peaks; in this case the treatment is more complicated, but the behaviour at long times is the same.

\begin{figure}[!h]
	 \begin{center}
                \includegraphics[width=0.7\columnwidth]{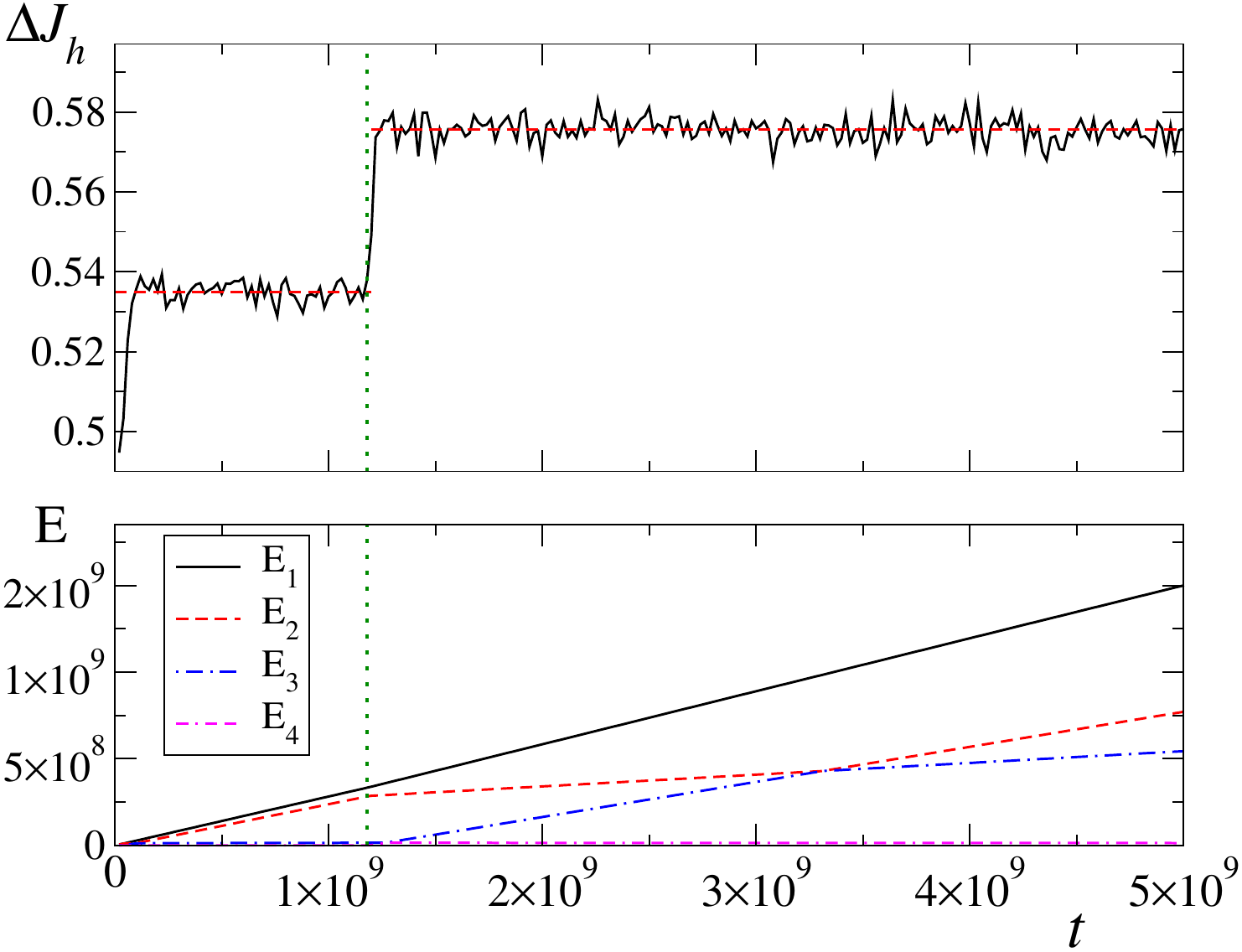} 
	 \end{center}
        \caption{
        Top panel: asymmetry of the energy flux $\Delta J_h$  vs time for a system of size $N=32$ connected to critical heat baths at $a_L= 1$, $a_R=5$.  The flux asymmetry at time $t$ is determined by averaging as in the previous figure.
        The asymmetry fluctuates around a value which corresponds to the average variation of energy of the peaks 
present in the system. 
	Bottom panel: peak energy vs time. The dashed lines in the
	top panel are obtained summing the growth rates of the peaks that are present at a given time. 
When the peak number increases, the asymmetry also increases, as indicated by the 
vertical dotted line.
}
	\label{fig:flusso_energia}
    \end{figure}

As for the asymmetry of the energy flux $\Delta J_h = J^{(L)}_h - J^{(R)}_h$, 
in Fig.~\ref{fig:flusso_energia} we plot its time dependence along with the
evolution of the energy contained in the various peaks.
The linear growth of the energy peaks confirms the expected constant growth rate.
Moreover, one can verify that the energy flux asymmetry coincides with the sum of peak growth rates: 
a jump in the current asymmetry is always related to the birth of a new peak.

\section{The metastable regime without localized states}
\label{sec.metastable}

In this section, we discuss the temporal regime preceding the birth of peaks.
Since metastability occurs only above the critical line, we consider a setup where both thermostats 
operate at infinite temperature, so that all the parametric curve connecting them is contained in the NT region.

So long as no peak is born, we can assume local equilibrium and make use of
the Onsager-matrix formalism (see Sec.~\ref{sec.model})
even though it has been developed with reference to the PT region.
In fact, as shown in~\cite{IP24}, thermodynamic properties smoothly vary while crossing the critical line,
the main difference being that fully stable states become only metastable.
In particular, when the thermostat parameters are close to one another, it is legitimate to invoke a perturbative approach and use a profile $\beta(m)$ valid in the proximity of the
critical curve $\beta=0$,\footnote{We go beyond the quadratic term found in~\cite{iubini23}
to have a better comparison with numerics, see below.}
\be
\beta(m) = \beta_0 + A(m-m_0)^2 + \frac{B}{m_0} (m-m_0)^3,
\label{eq.bm3}
\ee
where
\be
\label{eq:g}
 A = \frac{\overline L_{aa}(0)\overline L_{ah}(0)}{2 \det \overline L(0)} \equiv g \simeq 0.58,
\ee
\be
 B = \frac{\overline L_{aa}^2(0)\overline L_{ah}(0) \overline L_{hh}(0)}{3 (\det \overline L(0))^2}  \simeq 1.72,
\ee
with $\overline L$ referring to the scaled Onsager matrix defined in Eq.~(\ref{eq_OM}) evaluated at the
infinite-temperature point (see \ref{app.cubic}).
The parameters $\beta_0$ and $m_0$ are determined imposing
that the two thermostats operating at infinite temperature sit in 
$a_{L,R}=a_0(1\pm \Delta)$.

\begin{figure}[!ht]
\begin{center}
\includegraphics[width=0.7\textwidth,clip]{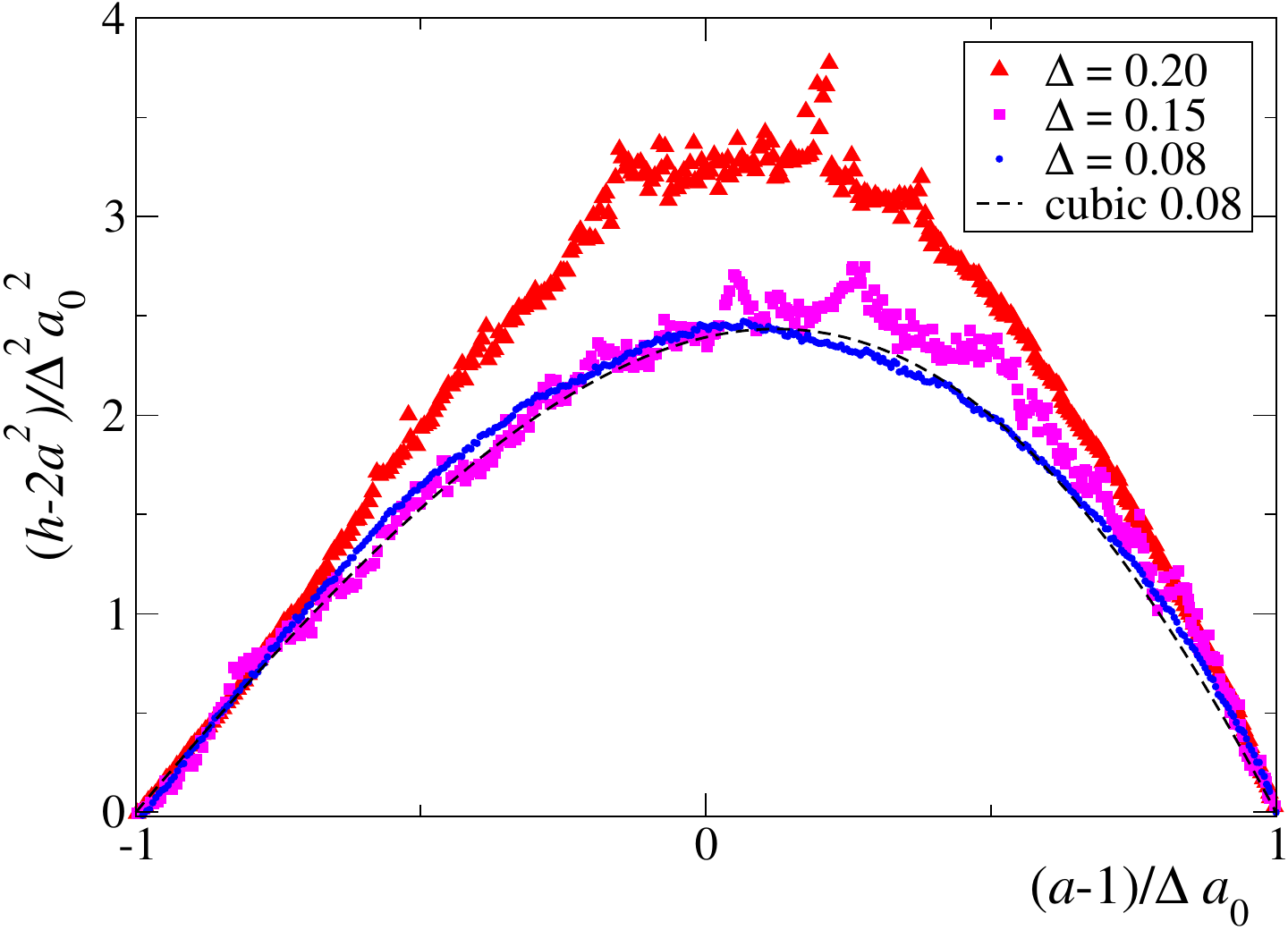}
\end{center}
\caption{Metastable path profiles for critical baths at mass densities
	$a_{L,R}=1\pm\Delta$, for $N=400$, and $\Delta=0.2, 0.15, 0.08$ (see the legend).
	Data have been obtained averaging over a time $2\times 10^9$ for $\Delta=0.20,0.15$
	and a time $10^{11}$ for $\Delta=0.08$.
}
\label{fig:prof_tot}
\end{figure}

Once we have a curve $\beta(m)$, we can use 
Eqs.~(\ref{mapping_a_mub},\ref{mapping_h_mub}) and the
approach used in Ref.~\cite{IP24} to obtain the parametric profile $h(a)$.
In Fig.~\ref{fig:prof_tot} we plot some metastable profiles using the rescaled
variables $x=(a-a_0)/(\Delta a_0)$ and $y=(h-2a^2)/(\Delta a_0)^2$.
We also show a fairly successful comparison of the numerical profile for the
smallest $\Delta=0.08$ with the profile predicted by the Onsager theory.
At the leading order, such theory gives a parabolic form, $y(x)=4g(1-x^2)$ ~\cite{iubini23}.

The clearly asymmetric shape justifies the cubic term in Eq.~(\ref{eq.bm3}).
We also remark that the negative concavity of the profile $y(x)$ is a manifestation
of Joule heating induced by the irreversible transport of energy and mass through the system~\cite{iubini23}.

\begin{figure}[!ht]
\begin{center}
\includegraphics[width=0.7\textwidth,clip]{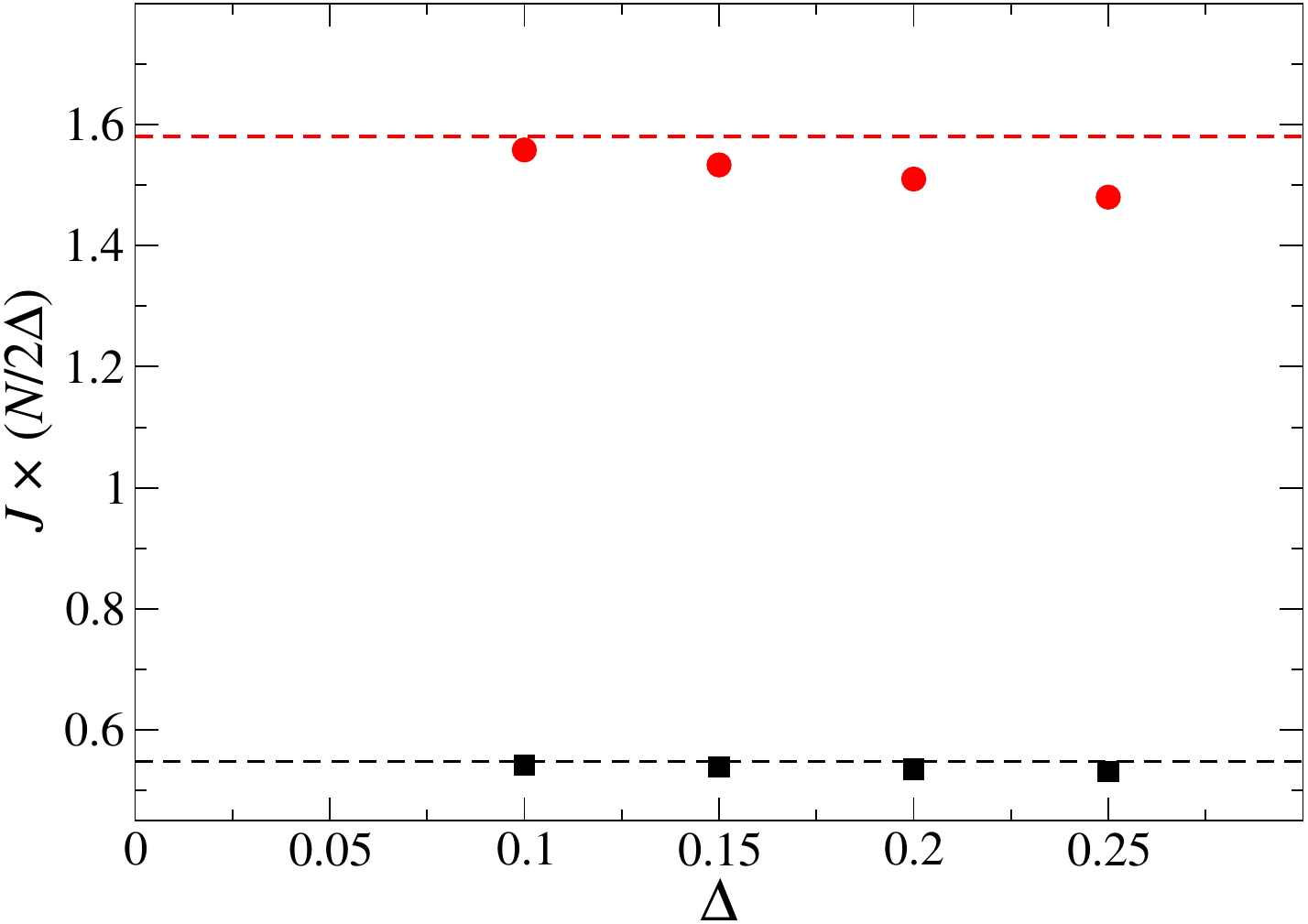}
\end{center}
\caption{Energy (red circles) and mass (black squares) stationary fluxes rescaled by the mass 
	gradient $2\Delta/N$ for a chain with $N=400$. 
	Symbols are the results of numerical simulations (running over times longer than $10^7$ time units), dashed lines
refer to the prediction from Onsager theory, see text. 
}
\label{fig:flux_tot}
\end{figure}

So long as localized states do not emerge, one can also use the 
Onsager formalism to determine the mass ($j_a$) and energy ($j_h$) fluxes.
As shown in Fig.~\ref{fig:flux_tot},
such metastable steady states are characterized by well defined currents $J_{a,h}$, both
proportional to the gradient $2\Delta/N$. As a result, transport is diffusive~\cite{Lepri_PhysRep}. 
The horizontal dashed lines correspond to the theoretical values determined via linear response theory at infinite
temperature. This can be done using Eqs.~(\ref{LCTeq1}-\ref{LCTeq2}) and the results obtained in 
Ref.~\cite{iubini23}. In practice, we obtain that the rescaled currents are
$|J_x| N/(2\Delta) =\overline{L}_{xa}(0)$, where $\overline{L}_{aa}(0)\simeq 0.548$
(lower line) and $\overline{L}_{ha}(0)\simeq 1.58$ (upper line).

It is worth stressing that such values are significantly smaller than those appearing in the steady states 
of the no-pinning case, where $|J_a| N/2\Delta = 4$ and $|J_h| N/2\Delta = 16$ ~\cite{GIP24}. This implies that pinned peaks, even when they are not 
large enough to grow indefinitely, contribute to an effective reduction of mass and energy transport.
  
Overall, the above observations account for a rather unusual pseudo-stationary transport involving self-generated NT states~\cite{baldovin21,Baldovin_2021_review,Iubini2017_Entropy}. 
The metastability of these states terminates with the emergence of growing peaks.
In the following, we address this phenomenon
by adopting two distinct approaches: (i) a dynamical one based on the behaviour of triplets of consecutive sites; 
(ii) a thermodynamic one, based again on the Onsager formalism.
So far, related phenomena of peak instability in thermodynamic setups have been studied either in 
isolated systems~\cite{rumpf07,rumpf09,rumpf21} or in open conditions imposed by an external reservoir~\cite{IP24}. The peculiarity of our setup is that explicit nonequilibrium (transport) conditions are now imposed by the presence of two distinct  boundary reservoirs.

\subsection{Dynamical approach}

Above the critical curve, localized states can appear and  are persistent, i.e. their growth is more likely than their reabsorption. This feature can be easily shown by considering a triplet of sites $(N=3)$ whose edges are thermalized at infinite temperature, but with different masses $a_L$ and $a_R$, so that the system is out-of-equilibrium.

\begin{figure} [!h]
\includegraphics[width=0.7\columnwidth]{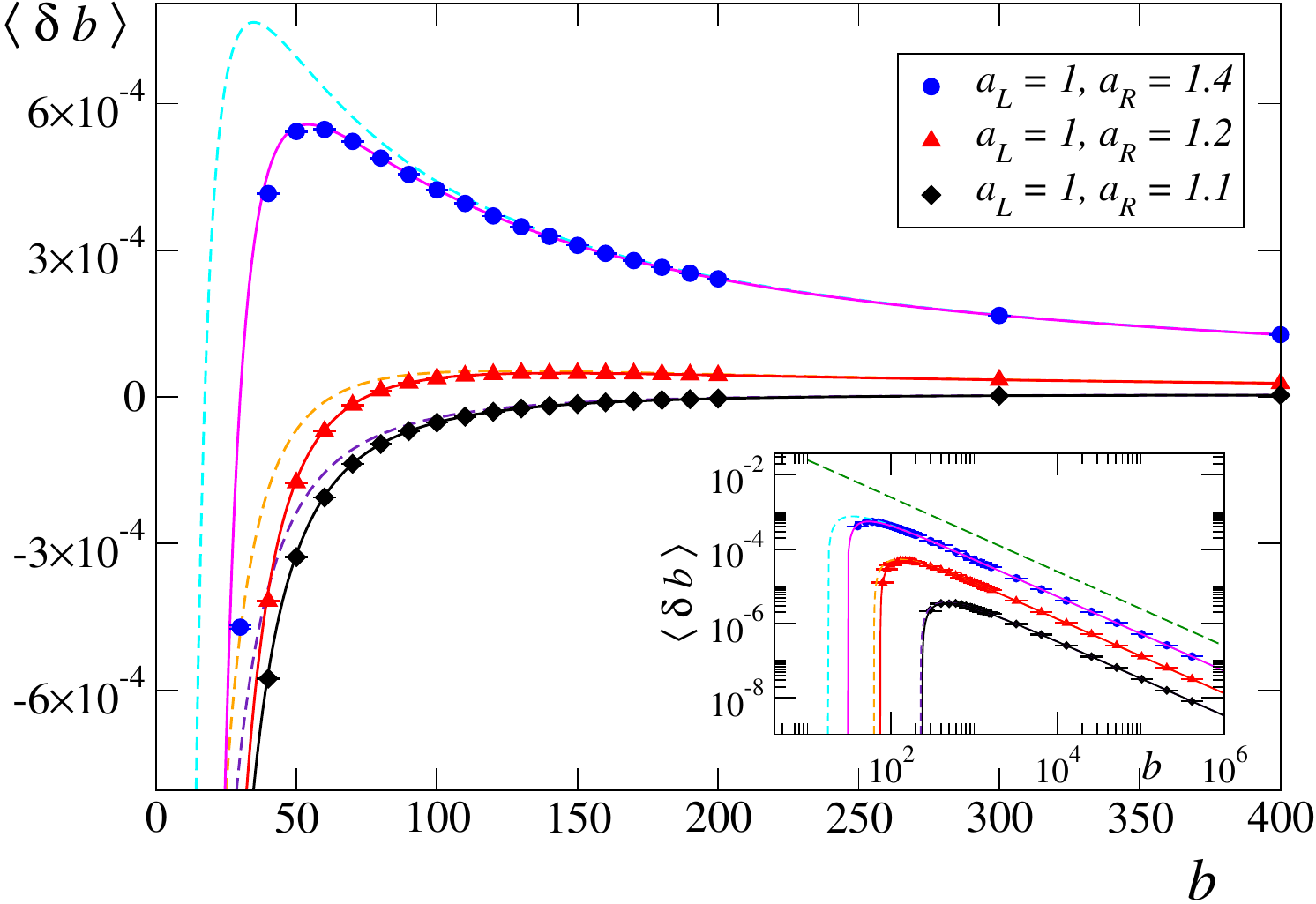} 
\caption{
Average mass variation $\langle \delta b \rangle$ per unit time of a peak with initial mass $b$ 
on the central site of a system of size $N=3$. 
The triplet is  connected to critical heat baths imposing masses $a_L$ and $a_R$, see legend. 
	Numerical averages (black diamonds, red up triangles, and blue circles) 
were obtained from $10^9$ independent unit-time updates of the triplet dynamics.
Lines refer to the analytic result in Eq.~(\ref{eq.vb}) and~\ref{app.triplet_b}:
dashed lines were obtained considering the expansion up to order $(1/b)^2$, continuous lines up to order $(1/b)^4$. Inset: same curves and data points in log-log scale. The green dotted line is a reference scaling $1/b$.
        }\label{MATHlinlinToy3}
    \end{figure}

In~\ref{app.triplet_b} we determine the average growth rate $\langle \delta b\rangle$
 of the central site of mass $b$, by solving the stochastic update dynamics of the triplet.
 In detail, $\langle \delta b\rangle$ is defined as the average mass variation of the central site with initial  mass $b$ after a
 time step. 
We explicitly find
\be
\langle \delta b\rangle = \frac{4}{3} \frac{a_0^2}{b} \Delta ^2 -\frac{2}{3} \frac{a_0^3}{b^2} \left(1-7 \Delta ^2\right) + {\cal O}\left(\frac{1}{b^3}\right),
\label{eq.vb}
\ee 
where $a_{R,L}=a_0(1\pm\Delta)$
and the explicit expression of the sub-leading terms, ${\cal O}(\cdots )$, is given 
in~\ref{app.triplet_b}.
This equation suggests that peaks grow with a rate $4a_0^2\Delta^2/3b$ if $b$ is large enough.
However, for $\Delta \ll 1$, there exists a critical value $b_c$ below which $\langle \delta b\rangle < 0$
shrinks, By equating the first two terms, we find $b_c=\frac{a_0}{2\Delta^2}$.
As expected, upon approaching equilibrium ($\Delta = 0$), $b_c$ diverges, i.e.  no persistent peaks appear.
In the opposite limit of increasing $\Delta$, which corresponds to a stronger gradient,
shorter and shorter peaks may grow; yet, sufficiently tall peaks still grow according
to Eq.~(\ref{eq.vb}). 
The analytical expression of $\langle \delta b\rangle$
fits well with simulation results, see Fig.~\ref{MATHlinlinToy3} and its caption for more details.

\subsection{Thermodynamic approach}
For $N>3$ the dynamics of the system quickly becomes rather awkward, so that
it is not treatable with the analytic methods employed in the previous subsection.
A complementary point of view on the problem consists in considering a very large system, ideally in the
thermodynamic limit, exchanging very small net currents with the external reservoirs.
In this limit, the assumption of local (quasi-)equilibrium is a reasonable hypothesis which allows to postulate 
the existence of smooth profiles $\beta(x)\leq 0$ and $m(x)$. 

At the lowest order in $\Delta\ll 1$, we can 
use Eq.~(\ref{eq.bm3}) up to the quadratic term, $\beta=\beta_0 + g(m-m_0)^2$, with critical thermostats in
$m_{1,2}=m_0(1\pm\Delta)$, which implies $\beta_0 =-gm_0^2\Delta^2$.

According to~\cite{IP24}, a local peak on $x$ becomes unstable when 
its mass overcomes the threshold $b_c^*=m(x)/2\beta(x)$. 
It is therefore necessary to identify the lowest threshold along the profile.
To leading order in $\Delta$, this condition is realized by the maximum of $|\beta(x)|$, 
that is $\beta_0$.
As a result, the instability threshold writes $b_c^*=1/(2g |m_0| \Delta^2)= a_0/(2g \Delta^2)$. 
Remarkably, this threshold displays the same scaling properties 
of its dynamical analogue for the triplet system, $b_c= a_0/(2 \Delta^2)$, but it explicitly depends on Onsager coefficients through $g$.
This residual discrepancy is not surprising if one considers the two approaches are conceptually very different.

\begin{figure}[t]
\begin{center}
\includegraphics[width=0.7\textwidth,clip]{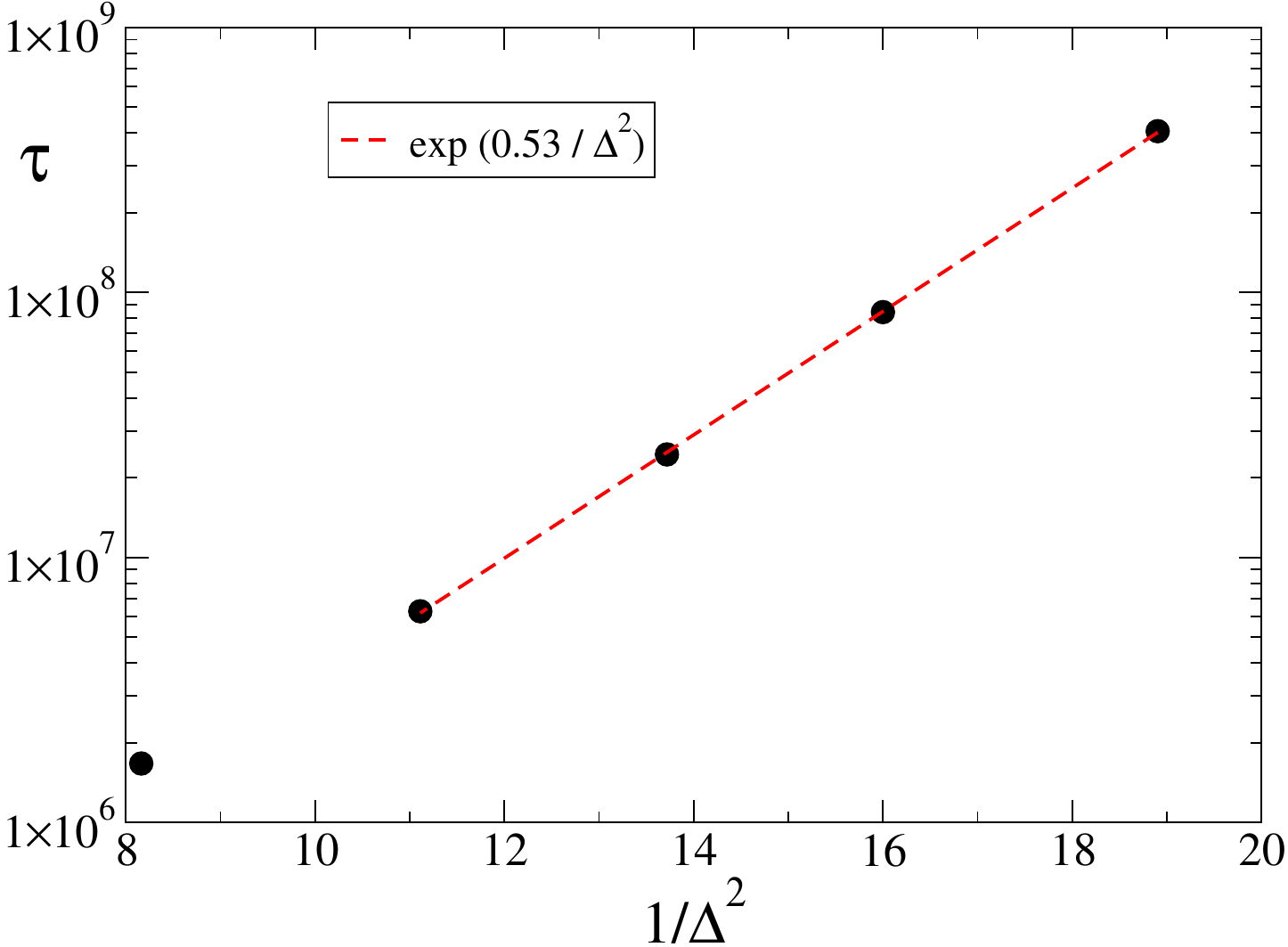}
\end{center}
\caption{
Average instability time $\tau$ of the nonequilibrium state with no peaks.  $\tau$ is computed as the average time needed to observe the 
first unstable peak starting from a homogeneous nonequilibrium profile. Simulations are performed for $N=400$ and averaged over $100$ 
independent realizations of the stochastic process. $N$ is large enough for the results to be independent of the chain length.
	Dashed line is a fit in the region of small $\Delta$, $\tau \simeq e^{0.53/\Delta^2}$. 
}
\label{fig:time_breather}
\end{figure}

Energetically, the crossing of the instability threshold corresponds to overcoming a potential barrier equal to 
$\Delta U=m^2/4|\beta|$~\cite{IP24}. With our assumptions, we obtain $\Delta U=(4g \Delta^2)^{-1}$.
This result allows obtaining an estimate of the typical lifetime $\tau$ of the metastable state through Kramers formula,
$\tau\simeq \exp(\Delta U)$. Explicitly, we obtain
\begin{equation} 
\label{eq.taudelta}
  \tau(\Delta) \simeq \tau_0 \exp \left( \frac{1}{4g \Delta^2} \right) \,.
\end{equation}
Numerical simulations performed on the metastable nonequilibrium state of relatively large systems, confirm that the 
average birth time of the first persistent peak scales exponentially with $\Delta^{-2}$, see Fig.~\ref{fig:time_breather}.
It is noteworthy that $\tau$ increases extremely rapidly for small $\Delta$, thus making 
the metastable regime effectively stable.
Finally, concerning the coefficient $\lambda$ of exponential growth $\tau\simeq \exp(\lambda \Delta^{-2})$, we numerically obtain $0.53$ (see Fig.~\ref{fig:time_breather}) 
to be compared with $(4g)^{-1}\simeq 0.43$.  One possible source of this discrepancy might be the indirect 
dependence of the prefactor $\tau_0$ on $\beta$ and therefore on $\Delta$,
as found in~\cite{IP24}. 


\section{Long-time configurations}
\label{sec.long-time}

The goal of this section is to investigate the long-time configurations when the whole system, or just a part of it, enters the NT region: the former case occurs in the presence of critical baths; the latter one may occur in the presence of positive$-T$ baths. 

Figures \ref{fig:ah_profile} and \ref{fig:aDh_profile} show that peaks form spontaneously and their number $N_b$ increases over time. However, on realistic time scales we have never observed an extensive number of peaks, because $N_b \ll N$ (or, more rigorously, $N_b/N$ decreases with $N$). This behaviour is a consequence of the super-exponential growth of the birth-time 
of a new peak as a function of the distance $\Delta$ between two continguous critical baths 
shown in Fig.~\ref{fig:time_breather}.
In fact, even when $a_L$ and $a_R$ are sufficiently far apart (i.e., $\Delta$ is not small), 
the appearance of new peaks splits the system into progressively smaller (and almost disconnected) subchains, separated
by the previous peaks which act as critical baths.
Hence, since the evolution slows down, the system effectively freezes 
in a configuration with just a few peaks. 

The overall growth can be roughly estimated with the help of
Eq.~(\ref{eq.taudelta}) by assuming that, starting from a given value $\Delta_0$, 
the formation of $N_b$ peaks uniformly partitions the interval so that
$\Delta(N_b) = \Delta_0/N_b$. 
Therefore, on average, the time necessary to observe the emergence of $N_b$ peaks is
\begin{equation}
   t(N_b) \simeq \tau_0  e^{cN_b^2/\Delta_0^2},
\end{equation}
from which
\begin{equation}
    N_b \sim \sqrt{\ln t} \; .
\end{equation}
The extremely slow growth of the number of peaks is attested by simulations, 
that have never allowed us to attain a number $N_b(t)$, of the order of the system size, $N$.

\begin{figure} [!ht]
                \includegraphics[width=0.7\columnwidth]{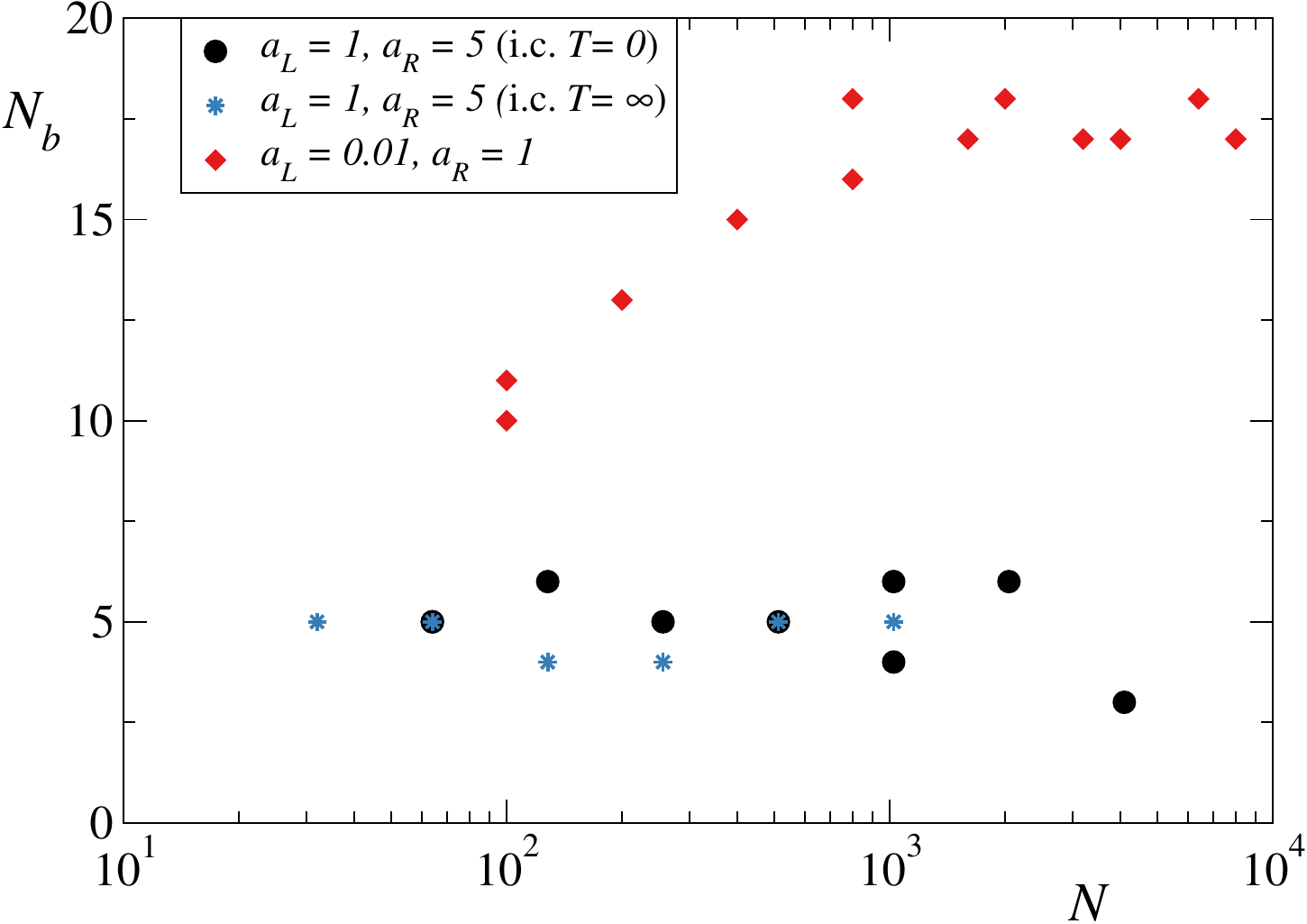} 
        \caption{
        Number of peaks at the end of the simulations with critical heat baths for systems with different sizes. For the simulations with $a_L = 1, a_R = 5$, 
	initial conditions (i.c.) were considered both with a linear spatial profile (black dots),  corresponding
	to $T=0$, and a profile generated by exponential distributions (blue stars), corresponding to $T=\infty$. 
	For the simulations with $a_L = 0.01, a_R = 1$ (red diamonds), 
	we considered only the infinite temperature initial condition. 
	Simulation times for the exponential initial conditions are at least $10^{10}$ units;
	simulation times for the linear initial conditions are at least $10^6$ units.
        }
        \label{fig.Nb}
    \end{figure}

Once $a_{L,R}$ are fixed, we expect $N_b$ to depend only weakly on $N$,
because the (maximal) simulation time cannot be varied by orders of magnitude.
 This is confirmed in
Fig.~\ref{fig.Nb} where we plot the ``final'' number of peaks, $N_b(N)$, for different setups:
different pairs of heat baths and different initial configurations for the same heat baths. 
There, we see that, for large $N$, $N_b$ is approximately constant and does not depend on the initial conditions, 
while it depends (as expected) on the separation $\Delta_0$ between the heat baths. 

\begin{figure} [!ht]
                \includegraphics[width=0.7\columnwidth]{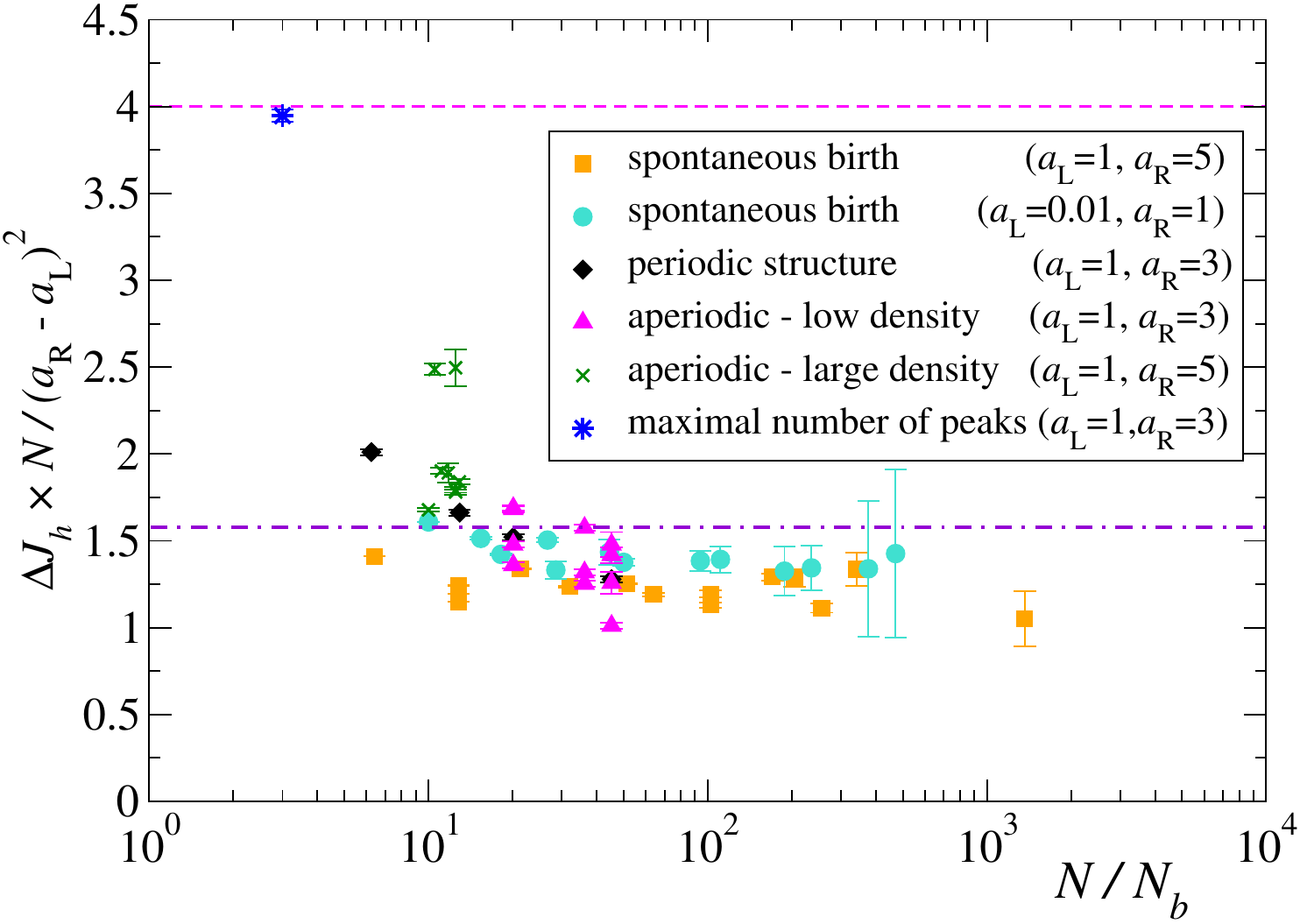} 
        \caption{
		Reduced asymmetry of the energy current as a function of
		the average distance between peaks, for
		different types of configurations.
		Dashed line $(y=4)$: periodic configuration with
		the maximal number of allowed peaks;
		dot-dashed line $(y=\bar L_{ah}\simeq 1.58)$: macroscopic theory, see~\ref{app.onsager}.
%
Orange squares and cyan circles are obtained from homogeneous initial conditions (spontaneous birth of localized states)
and show the regime of large and low average masses, respectively.
Black diamonds refer to periodic configurations of infinite-energy peaks with spatial period 
$p= 6,12,18, 36$ initialized on infinite-temperature background.
Magenta triangles and green crosses refer to disordered configurations of infinite-energy peaks
in the regime of low and large peak densities, respectively.
The blue star shows the state with maximal density of peaks.
All simulations are obtained with $100 \le N \le 8000$ and simulation times
larger than $10^6$, up to $10^{11}$.
Trajectories displaying peaks with negative growth rate are discarded. When $N_b$ varies during the simulation,
$\Delta J_h$ is obtained by averaging in the time
interval from the appearance of the last peak to the end of
the simulation.
        }
	\label{fig.Djh}
    \end{figure}

An important property of the asymptotic configuration is the asymmetry $\Delta J_h$ 
between the incoming and the outgoing energy current (\ref{eq:as_ener_curr}),
related to the energy drained by peaks.
In Fig.~\ref{fig.Djh} we plot $\Delta J_h N/(a_R-a_L)^2$ \textit{vs} the
average distance between consecutive peaks, $N/N_b$, for different types of configurations:
configurations where peaks appear spontaneously, starting from a homogeneous state
(orange squares and cyan circles);
periodic (black diamonds) and aperiodic (magenta triangles and green crosses) structures
of imposed peaks; see the legend for more details.
The scaling of the vertical axis has been suggested by two theoretical results (horizontal lines),
while the choice of the horizontal axis allows distinguishing between
configurations where $N_b$ scales with $N$ from those where $N_b$ is independent of $N$.

The upper dashed line corresponds to the reduced current for
the configuration with the \textit{maximal} number of peaks, consisting in one peak every third 
site.\footnote{If two peaks are closer, they may fall within the same triplet, thereby leading to
a strong redistribution of their masses and energies and  a fast dismantlement of the peaks themselves, see~\ref{app.triplet}.}
Such a periodic state is studied in~\ref{app.maxb}, where we show that it is stable
(i.e., all imposed peaks have a positive growth velocity) 
and where we determine analytically the parametric configuration of low-density sites: 
in the large$-N$ limit, such sites are located on the critical curve. 
This is a different way of confirming that the peaks impose infinite temperatures
in their surroudings.
Equation~(\ref{eq.Deltajh}) reveals $\Delta J_h$ 
is proportional to the gradient square of the mass, $(a_R-a_L)^2$, and 
is inversely proportional to the system size, $N$. More precisely,  the rescaled asymmetry 
is asymptotically equal to 4. 
Numerical simulations (see the asterisk) confirm the expected value, the tiny disagreement
being due to finite size effects.

The lower dotted-dashed line corresponds to
an alternative approach, based on the Onsager matrix formalism and developed in \ref{app.cp},
where we assume that the whole profile lies on the critical line and thereby solve 
the underlying differential equations for thermodynamic forces, Eqs.~(\ref{LCTeq1}-\ref{LCTeq2}).
The final expression of $\Delta J_h$ given in Eq.~(\ref{eq.Djh2}) is again proportional to 
$(a_L-a_R)^2$.
At variance with the dynamical result, the proportionality
constant now is $\overline L_{ah} \approx 1.58$: this value corresponds to the lower dashed
line in Fig.~\ref{fig.Djh}. 
This result corresponds to the limit of an arbitrarily large number of
peaks in a system where the limit $N\to\infty$ has been already taken. 

The most important feature emerging from these results is that the reduced current asymmetry
varies very little, at most a factor of three, for a large variety of
configurations whose density of peaks varies greatly, by three orders of magnitude.

\comment{
Another important feature of the localization process is the spatial distribution of peaks. 
In order to gain insights into this question, we can use
Eq.~(\ref{eq.taudelta})  to implement a simplified deterministic model where 
a new peak appears at the center of an interval delimited by an existing peak or by one
of the two reservoirs. Time (\ref{eq.taudelta}) starts from the birth of the new interval.
If we want to compare (as we will do) the results of this model with a simulation in which 
peaks arise spontaneously, we cannot think to compare \textit{spatial distributions},
because of the extreme slowness of the formation of peaks. We will therefore limit to
divide the system into three contiguous intervals $I_{1,2,3}$, and to count the fraction of
peaks present in each interval, $(n_1,n_2,n_3)$. We did that for a couple of different setups:
in the first setup, $a_L=1,a_R=5,N=2048$; in the second setup, $a_L=0.01,a_R=1,N=400$.
In the former case, the fractions relative to the formation of the first six peaks are
exactly the same: $(\frac{1}{2},\frac{1}{3},\frac{1}{6})$; in the latter case, the fractions
relative to the formation of the first fifteen peaks are only very slightly different:
$(\frac{2}{3},\frac{4}{15},\frac{1}{15})$ in the simulation,
$(\frac{11}{15},\frac{1}{5},\frac{1}{15})$ in the simplified model.
This comparison shows that when several peaks are present, their distribution is mainly
determined by the value of $\Delta=(a_2 -a_1)/(a_2+a_1)$ of each interval $(a_1,a_2)$ where
a peak will arise.
}

\section{Paths crossing the infinite-temperature line}
\label{sec.PNpaths}

In this section, we return to the original setup, where both thermostats operate inside the
positive temperature region and yet, the connecting parametric path ``wants'' to cross the critical line.
As sketched in Fig.~\ref{fig:sketch}c and argued in Sec.~\ref{sec.phenomenology},
in \ref{app.maxb}, and in \ref{app.onsager}, 
a parametric path connecting ${\bf P}_L$ and ${\bf P}_R$, initially crossing the critical line
is eventually composed ot three segements
${\bf P}_L{\bf Q}_L$, ${\bf Q}_L {\bf Q}_R$, and ${\bf Q}_R {\bf P}_R$, 
where the first and the last segment connect the thermostats to the critical line, while the middle one lies along
the critical line.
The question is where are ${\bf Q}_L$ and ${\bf Q}_R$ located?

As shown in Eq.~(\ref{eq:onsager2}), the parametric path departing from a given point in the positive-temperature region
is entirely determined by the ratio between energy and mass fluxes.
This must be true also close to the critical line in the vicinity of ${\bf Q}_L$ (${\bf Q}_R$) 
since we have not identified any singularity in the dependence of the Onsager matrices on $\beta$ (and $m$).
Hence the slope of the two paths joining in ${\bf Q}_L$ (${\bf Q}_R$) must be the same;
consequently, ${\bf P}_L{\bf Q}_L$ (${\bf P}_R{\bf Q}_R$) must end tangentially to the critical line. 
This condition alone suffices to identify a unique path connecting ${\bf P}_L$ (${\bf P}_R$) to the critical line.

Using the Onsager theory discussed in Sec.~\ref{sec.model}, it is possible to determine
the parametric profiles connecting ${\bf P}_L$ (${\bf P}_R$) to ${\bf Q}_L$ (${\bf Q}_R$),
finding that ${\bf Q}_{L,R}$ is a function of ${\bf P}_{L,R}$ only.
In fact, starting from ${\bf P}_L$ (${\bf P}_R$) there is a continuum of paths having a point in
common with the critical line, but only one of them is tangent to it.\footnote{More precisely,
there are two, because ${\bf Q}_L$ (${\bf Q}_R$) can be at the right or at the left of
${\bf P}_L$ (${\bf P}_R$).}
Therefore, returning to Fig.~\ref{fig:sketch}c, the shapes of the two ending paths are independent of one another: 
the very same ${\bf P}_L{\bf Q}_L$ can be combined with different ${\bf Q}_R{\bf P}_R$ segments 
to form a meaningful non-equilibrium stationary path 
(and the same is true when ${\bf P}_R$ is fixed and ${\bf P}_L$ varied).

\begin{figure} [!hb]
	\includegraphics[width=0.7\columnwidth]{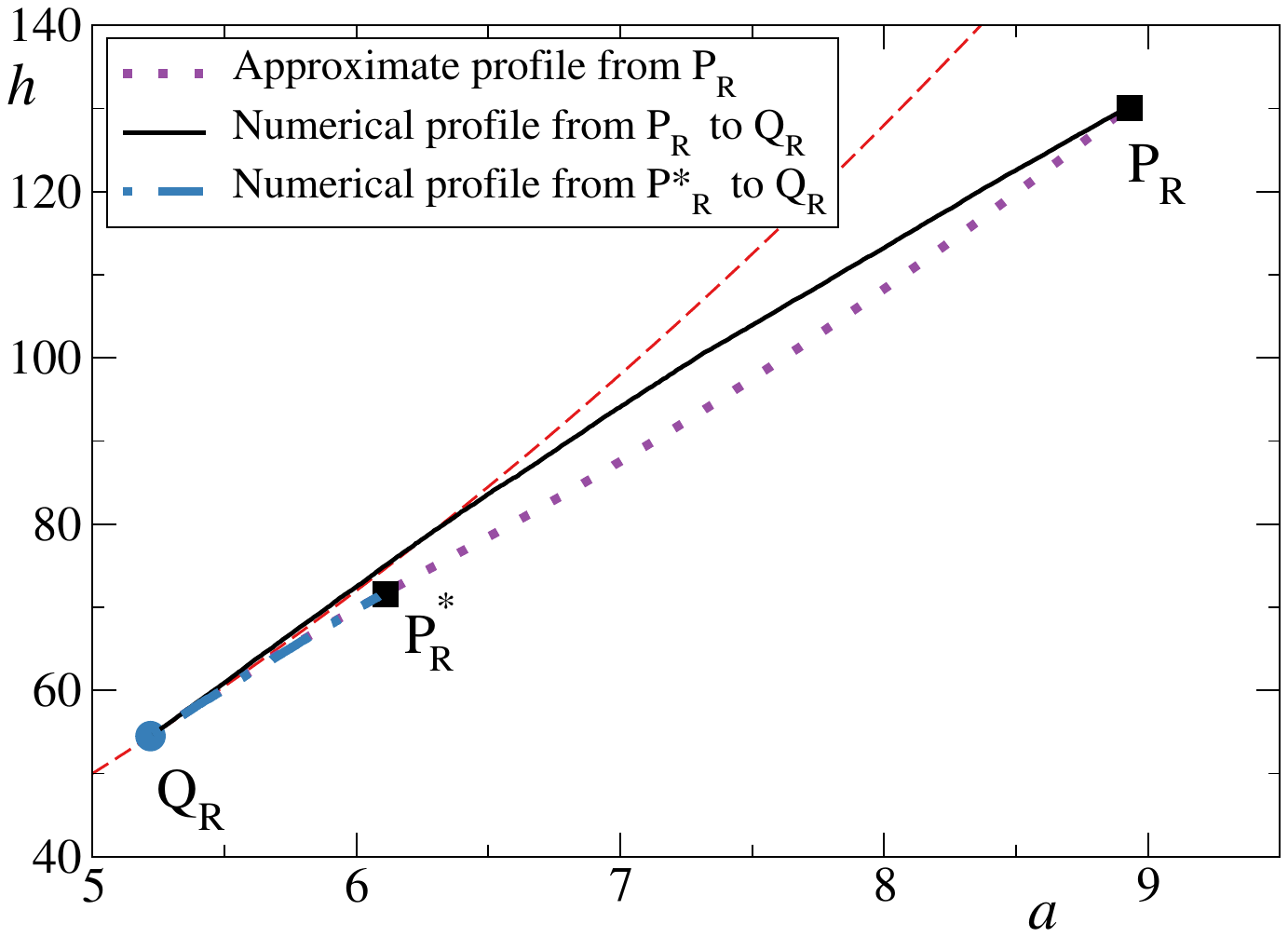}
        \caption{
	The purple dotted line is the analytical profile starting in ${\bf P}_R$ and 
forcing it to be tangent to the critical curve: it results from the Onsager theory
with the approximation $\overline{L}_{ij}(w) = \overline{L}_{ij}(0)$.
The black full (blue dotted-dashed) line is the  numerical profile with baths in ${\bf P}_R$
(${\bf P}^*_R$) and ${\bf Q}_R$.
The dashed red line is the critical curve.
        }
	\label{fig.Op}
    \end{figure}

The exact determination of the parametric profiles using the Onsager theory would require the knowledge
of the Onsager coefficients, which we do not have.
However, in proximity of the critical line we can assume that the reduced Onsager coefficients are constant,
$\bar L_{ij}(w)=\bar L_{ij}(0)$. 
Under this hypothesis, in \ref{app.Op} we find the analytical expression of the parametric profile
and in Fig.~\ref{fig.Op} we compare it to ``real'' (numerical) profiles.
If the bath ${\bf P}_R$ is not close enough to the critical line,
the real path is fairly different from the one obtained integrating
the Onsager equations in the hypothesis of constant coefficients.
However, starting from ${\bf P}^*_R$, the agreement is much better and the two curves are 
practically superimposed on each other.

When heat baths are chosen in such a way that the system crosses the critical line, the profile
is eventually  divided into three pieces, the two external ones being in the PT region
and tangent to the critical line.
We now determine the physical length of the three segments,
exploiting the fact that the mass flux is constant all along the whole path.

By denoting with $\delta_L$ and $\delta_R$ the yet unknown fractional length of the two ending segments, we can write
\begin{equation}
\frac{j_a(L)}{\delta_L} = -\frac{\overline{L}_{aa}(a_R-a_L)}{1-\delta_L-\delta_R} = \frac{j_a(R)}{\delta_R} 
\label{eq:12}
\end{equation}
where numerators represent the fluxes in the three segments, assuming each of them to be of unit length.
In particular, $j_a(L,R)$ can be determined solving Eq.~(\ref{LCTeq1}), while the flux in the middle segment
arises from the Onsager theory developed 
in~\ref{app.maxb}, with $a_L$ and $a_R$ denoting the mass density in ${\bf Q}_L$ and ${\bf Q}_R$, respectively.
Dividing such fluxes by the relative length of each segment, the three mass currents must be equal to each other.

In Eq.~(\ref{eq:12}) we have therefore two independent relations with two unknowns, $\delta_L$ and $\delta_R$,
which can be determined, obtaining
\begin{equation}
\delta_L = \frac{j_a(L)}{\tilde j_a}, \;
\delta_R = \frac{j_a(R)}{\tilde j_a}, \qquad \mbox{with} \qquad
\tilde j_a = j_a(L) + j_a(R) -\overline{L}_{aa}(a_R-a_L) .
\end{equation}
Hence we see that while the two ending segments have parametric profiles that are independent of the opposite
thermostat properties, their physical length encodes the information on the whole setup.

\section{Discussion and summary}
\label{sec.discussion}

The peculiar traits of the phenomenology discussed in this paper are consequence of two properties 
of the setup/model.
On the one hand, this is a boundary driven, out-of-equilibrium setup where the ends of
a chain are attached to two different heat baths. Since the model has two conserved quantities 
(the mass density $a$ and the energy density $h$),
it is a problem of coupled transport that can be solved by Onsager theory if local equilibrium is valid.
The parametric curve obtained by plotting the local energy as a function of the local mass is
an analytic curve within the positive temperature region.
On the other hand, the model, at equilibrium, displays a critical line $h_c(a)$ above which 
a finite fraction of the whole energy is localized in a few peaks (just one, in the thermodynamic limit).
These peaks represent spatial singularities which can also appear in the
out-of-equilibrium setup, even if both heat baths operate \textit{below} the critical line
in the well defined positive temperature region.

What happens in this case depends on the details of the dynamics, because the energy of a peak cannot
be constant in the out-of-equilibrum setup, since peaks intercept and store part of the transported energy,
thus leading to the growth of the peaks themselves. 
In Ref.~\cite{GIP24} a variant of the model where peaks can freely diffuse was studied.
Since the peaks can dissipate their increasing energy when they come in contact with the heat baths,
a NESS is attained whose parametric profile is perfectly analytic across the critical line (actually,
it is a linear profile).

In this paper, we studied a variant where peak diffusion is suppressed:
a feature which makes the stochastic system more similar to the DNLS model and the resulting
peaks more similar to the localized excitations (discrete breathers~\cite{flach08}) appearing in such 
Hamiltonian model~\cite{Campbell86,Rumpf04_pin}.
If peaks cannot diffuse, nothing prevents the energy contained in the peaks to increase forever.


At short times the system is in a metastable state because peaks have not yet appeared and the
profile is analytic even if it crosses the critical line; this regime
can actually last very long if the heat baths $X_L$, $X_R$ are mutually close.
Over sufficiently long times, peaks appear and act as infinite-temperature thermostats,
therefore inducing the parametric curve in the NT region to bend towards the critical line.
While the number of peaks increases, all other sites assume energy and mass progressively closer to 
the critical energy. 

The asymptotic state can be understood thinking of the chain as a sequence of three
subchains, the two external being in contact with the heat baths and ending tangentially on the 
critical line.
This latter feature follows from the continuity of the currents across the system. 
On the other hand, the non-analiticity of the energy current across the critical line
implies the non-analyticity of the nonequilibrium profiles. Physically, this stems from the fact that  
localized states can only appear in the inner subchain (above the critical line), from which they cannot ``escape'' 
 due to the pinning property. 
%
Therefore, the system is actually
phase-separated between homogeneous regions at positive temperature and an internal 
region hosting peaks of increasing energy.
In turn, the latter is composed of immobile peaks and sites with finite average masses
and energies, parametrically located on the critical line.

We conclude with a few comments about possible generalizations of our main result:
energy localization as a process to attain a (self-generated) infinite-temperature heat bath.
While here we have  mainly focused on the nonequilibrium setup, we remark that this property is also consistent 
with equilibrium in the condensed regime, where a single site hosts
the extra-energy while the rest of the system is an infinite-temperature background~\cite{Rumpf2004}.
The same is true, both at equilibrium and out-of-equilibrium, if the
energy is locally a function of the mass, $h_i=F(a_i)$, provided that $F(x)$ is convex~\cite{GIP24}.

It is also interesting to ask what happens if the physical dimension of the system increases.
First of all, at equilibrium, topology is completely irrelevant, but in the
out-of-equilibrium setup we must distinguish between the direction of mass/energy fluxes and
the transversal directions.
If energy peaks can diffuse, in~\cite{GIP24} it was proved that the overall picture is not
modified, only minor quantitative details intervene.
In the present case, where peaks are pinned, simulations of stripes (i.e., systems
of $N_c$ coupled chains of $N$ sites, with $N_c/N \ll 1$) 
suggest that the scenario is qualitatively similar to that one discussed in this paper.

Finally, as for possible extensions to Hamiltonian models displaying condensation transitions, like the DNLS 
equation~\cite{Rasmussen2000_PRL,johansson04}, 
we point out that the dynamic nature of the related localized states, discrete breathers, might introduce additional features
that have not been considered in this study. As an example, the dramatic dynamic decoupling of high-energy breathers from the surrounding lattice sites observed in~\cite{PRL_DNLS,IP24} is likely to act as a slowing down mechanism 
of the corresponding infinite-temperature heat baths. Implications for nonequilibrium settings~\cite{Iubini2017_Entropy} 
are not obvious and will require additional future work.



\ack
SI and PP acknowledge support from the MUR PRIN2022 project
“Breakdown of ergodicity in classical and quantum many-body systems”
(BECQuMB) Grant No. 20222BHC9Z.
\appendix

\section{Triplet evolution}
\label{app.triplet}

\subsection{The triplet update rule}
\label{app.triplet_rule}

Let us consider a mass triplet $\vec{I} = (c_1, c_2, c_3)$ that evolves to $\vec{F} = (c_1', c_2', c_3')$. The stochastic update of the triplet conserves the total mass  $\tilde{M}$ and the total energy  $\tilde{E}$,

\begin{equation}
    \begin{cases}
        \tilde{M} = c_1 + c_2 + c_3\\
        \tilde{E} = c_1^2 + c_2^2 + c_3^2
    \end{cases} .
\end{equation}
The mass triplet $\vec{I}$ can be viewed as a point in a three-dimensional space, the two constraints corresponding 
to the equation of a plane and the surface of a sphere, respectively. 
The new triplet  $\vec{F}$ must be chosen in the intersection of the two surfaces, that is a circumference which contains 
$\vec{I}$. 
Since the masses are positive real quantities, $c_1, c_2, c_3 \ge 0 $, the intersection is a full circumference only when $\tilde{E}\le \tilde{M}^2 / 2$, while it is composed of three disjoint arcs if $\tilde{E} > \tilde{M}^2 / 2$. In both cases, we can identify a new triplet $\vec{F}$ considering a rotation matrix $R(\theta)$ around the axis perpendicular to the plane and passing through the center of the circumference. The matrix $R(\theta)$ that rotates counterclockwise around axis $v = (1,1,1)/\sqrt{3}$ by an angle $\theta$ can be written as 
\begin{equation}
R(\theta) = 
 \begin{pmatrix}
 d & m & p\\
 p & d & m\\
 m & p & d
 \end{pmatrix}
\end{equation}
with
\begin{equation}
    \label{eq:App_defdmp}
    \begin{cases}
      d = \frac{1}{3} + \frac{2}{3} \cos\theta \\
      m = \frac{1}{3} (1- \cos \theta) - \frac{1}{\sqrt{3}} \sin \theta \\
      p = \frac{1}{3} (1- \cos \theta) + \frac{1}{\sqrt{3}} \sin \theta 
   \end{cases} .
\end{equation}

If the intersection is a full circumference, we obtain $\vec{F}$ by extracting the angle $\theta$ from the uniform distribution in the interval $[0, 2\pi]$ and then by rotating $\vec{I}$ with $R(\theta)$, $\vec{F} = R(\theta) \vec{I}$.
If, instead, the triplet $\vec{I}$ belongs to one of three disjoint arcs, there are two ways to update the mass triplet: 
we can either choose to evolve the triplet in any of the three arcs, or we can restrict $\vec{F}$ only to the points 
of the same arc containing $\vec{I}$. The first choice allows for a peak to diffuse and we refer to this rule as dynamics without pinning: 
the resulting dynamics has been studied in Ref.~\cite{GIP24}.

In this manuscript we consider the second choice, referred to as dynamics with pinning, in which the site of 
a triplet with large mass does not change in the update.    
Let us suppose that $\vec{I}$ belongs to an arc with $c_2 \gg c_1, c_3$. The points at the ends of the arc are 
\begin{equation}
    \vec{P_i} = (c_m, c_p, 0), \quad  \quad \vec{P_f} = (0, c_p, c_m)     
\end{equation}
with 
\begin{equation}
    \label{eq:App_cmcp}
    c_m = \frac{1}{2 } \left( \tilde{M} - \sqrt{2 \tilde{E} - \tilde{M}^2 }\right),  \quad \quad c_p = \frac{1}{2 } \left( \tilde{M} + \sqrt{2 \tilde{E} - \tilde{M}^2 }\right) . 
\end{equation}

The new triplet $\vec{F}$ can be obtained from a rotation of the point $\vec{P_i}$ using $R(\theta)$,
\begin{equation}
    \label{eq:App_tripl1}
    \vec{F} = R(\theta) \vec{P_i},  
\end{equation}
with $\theta$ generated from an uniform distribution in $[0, \alpha]$, where $\alpha$ is the angle subtended by the chord between $\vec{P_i}$ and $\vec{P_f}$. This angle $\alpha$ is given by the expression 

\begin{equation}
    \alpha = \arccos\left[\frac{1}{2} \left( \frac{\tilde{E} + \tilde{M}\sqrt{2 \tilde{E} - \tilde{M}^2} - 2 \tilde{M}^2/3 }{\tilde{E} - \tilde{M}^2/3}\right)\right] .
    \label{eq:App_alpha}
\end{equation}
Similarly, the same procedure can be applied to the other two arcs, by making a rotation from one of the end points of the respective arcs.

The described algorithm allows to satisfy both conservation laws. Moreover, using the uniform distribution to generate the angle $\theta$, given any initial triplet $\vec{I}$, every accessible final triplet $\vec{F}$ can be chosen with the same probability, so that the detailed balance condition is satisfied.

\subsection{Average mass variation of a peak}
\label{app.triplet_b}

Let us now consider the evolution of a peak of mass $b$ on the central site of a system composed of a single triplet $\vec{I} = (c_L, b, c_R)$. For the two external sites, the local masses $c_L$ and $c_R$ are random variables extracted from the mass distributions imposed by the heat baths. If we consider two reservoirs on the critical line with average mass densities $a_L$ and $a_R$, the mass distributions are 

\begin{equation}
       f_L(c_L) = \frac{1}{a_L} \exp\!\Big(\!\!- \frac{c_L}{a_L}\Big) \quad \quad \langle c_L \rangle = a_L, \quad \quad  \langle c_L^2 \rangle = 2a_L^2 
       \label{eq:_App_funzc1}
    \end{equation}
    \begin{equation}
       f_R(c_R) = \frac{1}{a_R} \exp\!\Big(\!\!- \frac{c_R}{a_R}\Big)  \quad \quad \langle c_R \rangle = a_R, \quad \quad  \langle c_R^2 \rangle = 2a_R^2.
       \label{eq:_App_funzc2}
    \end{equation}
Once $c_L$ and $c_R$ are extracted, $\vec{I}$ is updated to $\vec{F} = (c_L', b', c_R') = (c_L + \delta c_L, b + \delta b, c_R + \delta c_R) $ with the triplet move discussed in~\ref{app.triplet_rule}.
Since we want to study the conditions that lead to the growth of a peak, we will assume that $b$ is much greater than that of adjacent sites, $b \gg c_L, c_R$, that is a condition in which the new triplet $\vec{F}$ is determined by the expression (\ref{eq:App_tripl1}). The mass variation of the peak can therefore be written as
\begin{equation}
   \delta b = b' - b = ( \vec{F} - \vec{I} \, )_2  = p c_m + d c_p - b.
\end{equation}
The mass variation of the peak is a function of three random variables, $c_L$, $c_R$ and the angle $\theta$  in the $d, p, m$ components of the rotation matrix $R(\theta)$, see Eq.~(\ref{eq:App_defdmp}). It is also a function of the peak mass $b$, that is not extracted from a distribution since we want to know how it varies on average for a fixed value of $b$.
While $c_L$ and $c_R$ are generated by independent distributions $f_L$ and $f_R$, the angle $\theta$ has $\alpha$ as maximum amplitude, defined in Eq. (\ref{eq:App_alpha}),  which is a function of $c_L, c_R$ and $b$, i.e. $\alpha = \alpha(c_L, c_R, b)$. Once particular values of $c_L, c_R$ and $b$ are fixed, $\theta$ is a random variable uniformly distributed in the interval $[0, \alpha(c_L, c_R, b) ]$ and therefore it has a conditional distribution function:

\begin{equation}
    g(\theta| c_L, c_R, b) = \frac{1}{\alpha(c_L, c_R, b)} \Theta(\alpha - \theta) \Theta(\theta),
\end{equation}
where $\Theta$ is the Heaviside function. 
In our case the conditional distribution enters in the expression of the joint distribution function for the three random variables as follows,
\begin{equation}
    f(\theta, c_L, c_R, b ) = f_L(c_L) f_R(c_R) g(\theta| c_L, c_R, b).
\end{equation}
With this expression we can analytically derive the value of the average mass variation of a peak with mass $b$ in a time step,
\begin{equation}
  \begin{split}
      \langle \delta b \rangle &= \iiint dc_L dc_R d\theta  f(\theta, c_L,c_R, b) \delta b  \\
     &=  \int_0^\infty dc_L \int_0^\infty  dc_R f_L(c_L) f_R(c_R) \left[   \langle p \rangle_ \theta  c_m  + \langle d \rangle_ \theta  c_p -b \right],
  \end{split}
  \label{contodb1}
\end{equation}
where $c_m$ and $c_p$ are defined in Eq.~\eqref{eq:App_cmcp} and we denote with $\langle \cdot \rangle_\theta$ the average over the uniform distribution of $\theta$,
\begin{equation}
  \langle \cdot \rangle_\theta = \frac{1}{\alpha} \int_0^\alpha \cdot \quad d\theta    .
\end{equation}
By substituting the values of $d$ and $p$ from (\ref{eq:App_defdmp}) averaged over $\theta$ we obtain

\noindent\begin{equation}
\begin{split}
   \langle \delta b \rangle=  \int_0^\infty dc_L \int_0^\infty  dc_R f_L(c_L) f_R(c_R) &\left[  \frac{c_m}{3\alpha} \left( ( \alpha - \text{sin} \alpha) - \sqrt{3} (\text{cos} \alpha -1) \right) \right. \\
   &+ \left. \frac{c_p}{3\alpha}  ( \alpha + 2\text{sin} \alpha) - b  \right]. 
\end{split}
\end{equation}

We can now use the hypothesis that $b\gg c_R, c_L$ to perform a Taylor expansion of $\delta b$ in the powers of $1/b$. From the expansion of $c_m$, $c_p$ and $\alpha$ we can derive the average variation of $b$, here reported up to the order $1/b^4$:
\begin{equation}
\begin{split}
\noindent \langle\delta b \rangle \approx& \int_0^\infty \int_0^\infty dc_L dc_R f_L f_R \left[ \frac{1}{6 b} (c_L^2 - 4 c_L c_R + c_R^2) + \frac{1}{12 b^2} (c_L^3 - 5 c_L^2 c_R \right. \\ & - 5 c_L c_R^2  + c_R^3) 
+ \frac{1}{120 b^3} (c_L^4 - 26 c_L^3 c_R - 154 c_L^2 c_R^2  - 26 c_L c_R^3 + c_R^4) \\
&\left. -\frac{1}{240 b^4} (c_L + c_R)(7 c_L^4 +40 c_L^3 c_R + 486 c_L^2 c_R^2 + 40 c_L c_R^3 + 7 c_R^4)  \right].
\end{split}
\label{dm_nonmediato}
\end{equation}

Let us now average over the mass distributions $f_L$ and $f_R$. If we consider critical heat baths with the distributions (\ref{eq:_App_funzc1}) and (\ref{eq:_App_funzc2}), the expectation values of the powers $c^n$ of the local masses are 
\begin{equation}
    \langle c^n_i \rangle = a_i^n \, n!, \quad \quad i = L,R, 
\end{equation}
from which we can write 
\begin{equation}
\begin{split}
 \noindent \langle \delta b \rangle &=  \frac{(a_R - a_L)^2}{3b}  + \frac{1}{6b^2}(a_L + a_R)(3 a_L^2 - 8 a_L a_R + 3 a_R^2) +  \\
 &+ \frac{1}{30 b^3}(6 a_L^4 - 39 a_L^3 a_R - 154 a_L^2 a_R^2 - 39 a_L a_R^3 + 6 a_R^4) - \\
 &- \frac{1}{10 b^4} (a_L + a_R)(35 a_L^4 + 12 a_L^3 a_R + 251 a_L^2 a_R^2 + 12 a_L a_R^3 + 35 a_R^4) + O(\frac{1}{b^5}).  
\end{split}
\label{eq:App_db4}
\end{equation}

Finally, if we rewrite $a_L$ and $a_R$ as $a_{R,L}=a_0(1\pm\Delta)$,
Eq.~(\ref{eq:App_db4}) becomes 
\noindent\begin{equation}
  \begin{split}
    \langle \delta b \rangle =& \frac{4}{3} \frac{a_0^2}{b} \Delta ^2 -\frac{2}{3} \frac{a_0^3}{b^2} \left(1-7 \Delta ^2\right) -\frac{2}{15} \frac{a_0^4}{b^3} \left(55-95 \Delta ^2+16 \Delta ^4\right)\\
    &-\frac{1}{5} \frac{a_0^5}{b^4} \left(345-82 \Delta ^2+297 \Delta ^4\right)+O\left(\frac{1}{b^5}\right)  .
  \end{split} .
\end{equation}
From this expression we can see that if the two reservoirs impose an asymmetry in the system, $\Delta > 0$,  the first term of the expansion is positive and leads to the growth of a peak. However, the second and the following terms can be negative and can make the average mass variation negative when $b$ is not large enough to make them negligible. 
Hence, the sign of the average-mass variation depends on the competition between different terms;
a peak can emerge if the critical mass $b_c$, such that $\langle \delta b \rangle = 0$, is exceeded. 
An estimate of this threshold can be easily found by determining the mass $b$ such that the first two 
terms of the expansion compensate each other, i.e.
\begin{equation}
   b_c = \frac{a_0(1 + O(\Delta^2))}{2\Delta^2}.
\end{equation}

\section{Nonequilibrium ``steady state'' in systems with a finite density of peaks}
\label{app.maxb}

In this appendix we study analytically the properties of the state characterized by 
a finite density of peaks, using a discrete approach.
In the following appendix, we will use a continuum approach, based on the Onsager theory.

\subsection{Mass and energy profiles}
\label{app.mep}

With our stochastic dynamics two peaks, must be at the minimal distance of three lattice sites, otherwise they would interact with each other, leading to the  redistribution of their masses. The maximal number of peaks is therefore obtained by considering periodic peak configurations, with period 3.  
As a further constraint, peaks cannot be placed on the first and last sites of the chain as they would be in contact with the thermal reservoirs. 
For simplicity, we will assume $N$ to be  a multiple of 3, $N = 3k$ with $k>2$, 
and the peaks are located on the sites $3j-1$ ($j\in[1,k]$).
Peak masses are denoted with $b_j\equiv c_{3j -1}$. The sites to the right of a peak will have index $i$ that is a multiple of 3, i.e. $i = 3j$, while the sites to the left will have an as index $i = 3j -2$. 
In these configurations, every triplet of adjacent sites contains one and only one peak.
The masses and energies of the sites in the middle of two successive peaks 
stabilize at finite values $(a_i, h_i)$, which will be determined by the dynamic process and are linked to mass distributions that are not known a priori, as are those of the heat baths at the ends of the chain.    

Let us consider the mass $c_{3j}$ of one of the internal sites to the right of a peak, $i = 3j$ with $j=1,\dots , k-1$, which can be updated when one of the triplets $T_{3j-1} = (c_{3j -2}, b_j , c_{3j})$, $T_{3j} = (b_j, c_{3j}, c_{3j+1})$, or $T_{3j+1} = (c_{3j }, c_{3j+1}, b_{j+1})$ is chosen. These triplets are chosen on average once in every Monte Carlo step, so the total average variation of $c_{3j}$ is given by the sum of the averages made according to the three triplets,

\begin{equation}
    \langle \delta c_{3j}\rangle = \langle \delta c_{3j}(T_{3j-1})\rangle + \langle \delta c_{3j}(T_{3j})\rangle + \langle \delta c_{3j}(T_{3j+1})\rangle.
\end{equation}

Using results derived in \ref{app.triplet_rule}, we find
\begin{eqnarray}
	\langle \delta c_{3j}) \rangle_\theta &=& \frac{1}{2}(c_{3j-2} + 2c_{3j+1} - 3c_{3j}) + O(1/b_j, 1/b_{j+1}),
    \label{eq:App_dc3j}  \\ 
	\langle \delta c_{3j+1}) \rangle_\theta &=& \frac{1}{2}(2c_{3j} + c_{3j+3} - 3c_{3j}) + O(1/b_j, 1/b_{j+1}) ,
    \label{eq:App_dc3j+1}
\end{eqnarray}
where $\langle\cdots\rangle_\theta$ is the average over the distribution of the angle $\theta$.
Therefore, in the limit of infinitely high peaks, we obtain 
\begin{equation}
\begin{cases}
    \langle \delta c_{3j} \rangle_\infty = \frac{1}{2}(a_{3j - 2} + 2a_{3j + 1} - 3 a_{3j})\\
    \langle \delta c_{3j+1} \rangle_\infty = \frac{1}{2}(2a_{3j} + a_{3j + 3} - 3 a_{3j+1}).  
\end{cases}
\end{equation}
If we now assume that the subsystem including all $i$ non-peak sites reaches a nonequilibrium steady state, then the average masses $a_i$ must be 
such that the average mass variations $\langle \delta c_i \rangle$ vanish. 
By canceling the average variations of all internal sites, $i\neq 1 ,N$, we obtain the system of equations,
\begin{subequations}
   \begin{equation}
      a_{3j - 2} + 2a_{3j + 1} - 3 a_{3j} = 0 \quad \quad 1\le j \le k-1
    \label{eq:App_sl1a}
    \end{equation}
    \begin{equation}
      2a_{3j} + a_{3j + 3} - 3 a_{3j+1} = 0  \quad \quad 1\le j \le k-1  
      \label{eq:App_sl1b}
    \end{equation}
\label{eq:App_sl1}
\end{subequations}
which can be complemented by the boundary conditions for the sites in contact with heat baths, 
\begin{equation}
    a_1 = a_L, \quad a_N = a_R .
\end{equation}
The average masses $a_i$ can therefore be treated as $2k$ unknowns of a system of $2k$ coupled linear equations. The system (\ref{eq:App_sl1}) can be rewritten in a more regular version by summing Eqs.~(\ref{eq:App_sl1a}) and (\ref{eq:App_sl1b}) for the same $j$ and for consecutive $j$, obtaining

\begin{subequations}
   \begin{equation}
        a_{3j - 2} - a_{3j} - a_{3j +1} + a_{3j + 3} = 0 \quad \quad 1\le j \le k-1
    \label{eq:App_sl2a}
    \end{equation}   
    \begin{equation}
        a_{3j -3} - a_{3j-2} - a_{3j} +  a_{3j + 1} =  0  \quad \quad 1\le j \le k-1
      \label{eq:App_sl2b}
    \end{equation}
\label{eq:App_sl2}
\end{subequations}
Starting from Eq.~(\ref{eq:App_sl2b}) with $j = k-1 = N/3 - 1$, we can express $a_{N -5}$ as a combination of $a_{N -3}$, $a_{N - 2}$  and $a_N = a_R$, from which we can derive $a_{N -6}$ using Eq. (\ref{eq:App_sl2a}) with $j = k-1$ and so on. It is easily demonstrated by induction that the following relations hold,
\begin{equation}
    \begin{cases}
        a_{N  - 3m - 2} = m a_{N - 3} + a_{N -2} - m a_R & m \ge 1\\
        a_{N  - 3m } = m a_{N - 3} - (m-1) a_R  & m \ge 2
    \end{cases}
    \label{Eq:App_sl3}
\end{equation}
We can then use Eq. (\ref{eq:App_sl1b}) with $j = k-1$ to express 
\begin{equation}
    a_{N- 3} = \frac{1}{2} (3 a_{N-2} - a_R),
\end{equation}
which can be substituted in (\ref{Eq:App_sl3})  obtaining
\begin{subequations}
     \begin{equation}
      a_{N - 3m - 2} = \left( \frac{m}{2} + 1 \right) a_{N-2} - \frac{3m}{2} a_R \quad \quad m \ge 1
    \label{Eq:App_sl4a}
    \end{equation}
    \begin{equation}
       a_{N - 3m } = \frac{m}{2} a_{N-2} - \left(\frac{3m}{2}  - 1 \right) a_R \quad \quad  m \ge 2  
       \label{Eq:App_sl4b}
    \end{equation}
    \label{Eq:App_sl4}
\end{subequations}
If now we consider (\ref{Eq:App_sl4a}) with $m = k-1$
\begin{equation}
    a_L = a_1 = \frac{3k - 1}{2} a_{N-2} - \frac{3k -3}{2} a_R
\end{equation}
we can espress $a_{N-2}$ as a function of $a_L$ and $a_R$,
\begin{equation}
    a_{N-2} = \frac{2a_L + 3(k - 1) a_R}{3k -1}.
\end{equation}
Finally, by substituting $a_{N-2}$ in (\ref{Eq:App_sl4}), we obtain the expression for the average masses $a_i$ of the sites to the right and left of the peaks, i.e. with $i \bmod{3} \neq 2$, as a function of $a_L$ and $a_R$:
\begin{equation}
    a_i = a_L + \frac{i-1}{N-1} (a_R- a_L) ,\quad \quad 2\le i \le N-2,\quad i \bmod{3} \neq 1 .
    \label{eq:App_linprof}    
\end{equation}
Therefore, the solution is a linear spatial profile.

The same analytical approach for the average masses $a_i$ can be applied to derive a system of equations to obtain the average energies $h_i = \langle c_i^2\rangle$, 
\begin{equation}
\begin{cases}
    h_{3j-2} +2h_{3j+1} - 6h_{3j} + 2\langle c_{3j-2}c_{3j}\rangle + 4 \langle c_{3j}c_{3j +1} \rangle  = 0 &1\le j \le k-1\\
    2h_{3j} +h_{3j+3} - 6h_{3j+1} + 4\langle c_{3j}c_{ 3j +1} \rangle + 2\langle c_{3j+1}  c_{3j+3}\rangle = 0 &1\le j \le k-1 .
\end{cases}
\label{eq:App_sen1}
\end{equation}

In the extended system, the mass correlations $\langle c_i c_j \rangle$ are unknown and the system of equations (\ref{eq:App_sen1}) is not closed. 
In principle, the problem can be solved 
because the equations to determine them do not require higher-order correlations. The outcome
is a system of $N\times N$ coupled equations, rather than of $N$ equations as for the masses. 
We have obtained the explicit expression of such system, but
the identification of a closed-form solution for a generic size $N$ is a prohibitive task and we do not tackle it. However, by solving numerically the system, it is still possible to identify the dependence 
of $h_i =\langle c_i^2\rangle$ and $c(i,j)=\langle c_i c_j\rangle$ on the size $N$, for large $N$.
In particular, the spatial profile for the excess energies $\epsilon_i =h_i - 2a_i^2$ 
remains positive with a parabolic shape. The maximum value can be found in the center of the system, that is a site $i\sim N/2$, and vanishes as $1/N$ as the size increases.
If we instead keep the site indices $i$, $j$ fixed and vary only $N$, both $\epsilon_i$ and $c(i,j)$ decrease as $1/N^2$ as the size increases. 
In conclusion, in the thermodynamic limit the parametric profile $(a_i, h_i)$ flattens on the critical line, 
while the masses on different sites become uncorrelated. 

\subsection{The mass and energy currents}
\label{app.mec}


Following the notation introduced in Sec.~\ref{sec.model}, the mass flux on the left boundary is defined as


\begin{equation}
    J_a^{(L)} =  \langle J^{1,2}_a(T_2)\rangle = - \langle \delta c_1(T_2)\rangle =  \langle c_1 - c'_1 \rangle.
\end{equation} 
For the system with the maximal number of infinite-mass peaks, $\langle c'_1 \rangle = (a_L + a_3)/2$, therefore
\begin{equation}
     J_a^{(L)} =  a_L - \frac{a_L + a_3}{2} = - \frac{a_3 - a_L}{2}.
\end{equation}
By replacing the mass linear profile from Eq. (\ref{eq:App_linprof}) we obtain
\begin{equation}
     J_a^{(L)} =   -  \frac{a_R - a_L}{N-1}.
\end{equation}

Similarly, we can derive the mass flux at the right end of the chain 
\begin{equation}
    J_a^{(R)} =  \langle \delta c_N(T_{N-1})\rangle = - \frac{a_R - a_L}{N-1}.
\end{equation}
Since $J_a^{(L)} = J_a^{(R)}$, there is no mass flux asymmetry, $\Delta J_a = 0$.

As for the energy flux $J_h^{(L)}= - \langle \delta c_1^2(T_2)\rangle$, we find
\begin{equation}
        J_h^{(L)} = - \frac{4}{N-1} (a_R - a_L)a_L - \frac{8}{3}\frac{(a_R - a_L)^2}{(N-1)^2} - \frac{\epsilon_3}{3}.
\end{equation}
In the same way we can obtain right energy flux
\begin{equation}
    J_h^{(R)} =  -\frac{4}{N-1} (a_R - a_L)a_R + \frac{8}{3}\frac{(a_R - a_L)^2}{(N-1)^2} + \frac{\epsilon_{N-2}}{3}. 
\end{equation}

For both currents, the energies $\epsilon_3$, $\epsilon_{N-2}$  in excess of the critical values $2a_3^2$ and $2a_{N-2}^2$ contribute as terms of order $1/N^2$. Finally, by adding together $J_h^{(L)}$ and $J_h^{(R)}$ we obtain the analytical expression for leading term of the energy flux asymmetry $\Delta J_h$ 
\begin{equation}
    \Delta J_h = \frac{4}{N} (a_R - a_L)^2 + O(1/N^2).
	\label{eq.Deltajh}
\end{equation}

\section{Profiles within the Onsager theory}
\label{app.onsager}

With reference to the rescaled Onsager coefficients, any parametric profile (in the positive temperature region) 
satisfies the following differential equations
\begin{eqnarray}
j_a &=& -\frac{\overline{L}_{aa}}{m^2} \frac{dm}{dy} - \frac{\overline{L}_{ah}}{m^3}\frac{d\beta}{dy}  \nonumber \\
j_h &=& +\frac{\overline{L}_{ah}}{m^3} \frac{dm}{dy} + \frac{\overline{L}_{hh}}{m^4} \frac{d\beta}{dy}
\label{eq:onsager}
\end{eqnarray}
where the mass and energy profiles are both constant.
By taking the ratio between the two equations we find that
\begin{equation}
\frac{d\beta}{dm} = - \frac{\overline{L}_{aa}m^2+ (j_a/j_h)\overline{L}_{ah}m}
                      {(j_a/j_h)\overline{L}_{hh} + \overline{L}_{ah}m}
\label{eq:onsager2}
\end{equation}
This equation shows that, given an initial condition (fixed by a given selection
of the thermostat parameters), the parametric profile is uniquely determined by
the ratio between mass and energy profiles.

From now on, we focus on the critical region and neglect the dependence of
the rescaled Onsager coefficients on the inverse temperature $\beta$, with
the goal of determining analytical expressions of the associated profiles.
In the first subsession we determine the shape of a profile supposedly
converging to the critical line. In the second subsection we obtain a metastable profile
that enters in the NT region. The third subsection is devoted to
the characterization of a profile lying along the critical line.

\subsection{Convergence to the critical line}
\label{app.Op}

Assuming that the rescaled Onsager coefficients do not depend on $\beta$, Eq. (\ref{eq:onsager2}) is a separable differential equation.
Defining an initial condition for the parametric profile $\beta(m_0) = \beta_0$, we obtain

\begin{equation}
    \beta - \beta_0 = - \int_{m_0}^m dm \,  \frac{\overline{L}_{ah} m  + \overline{L}_{aa} m^2 \rho}{\overline{L}_{ah} m \rho   + \overline{L}_{hh}} ,
    \label{eq:onsager3}
\end{equation}
where $\rho = j_h/j_a$ and the rescaled Onsager coefficients are constants.
By solving the integral in the right-hand side of Eq. (\ref{eq:onsager3}) we obtain 
\begin{equation}
\begin{split}
    \beta(m, \rho) =\,& \beta_0 + \frac{m-m_0}{\rho} \left(\frac{\overline{L}_{aa} \overline{L}_{hh}}{\overline{L}_{ah}^2 } - 1 \right) - \frac{(m^2 - m_0^2)}{2 \overline{L}_{ah}} \\ &+\frac{\overline{L}_{hh} \overline{L}_{aa}^2
    - \overline{L}_{aa}\overline{L}_{hh}^2}{\overline{L}_{ah}^3\rho^2} \ln \left( \frac{\overline{L}_{hh} + \overline{L}_{ah}m\rho}{\overline{L}_{hh} + \overline{L}_{ah}m_0\rho}   \right).
    \label{eq:onsager4}
\end{split} 
\end{equation}
Different values of the ratio $\rho$ correspond to different parametric profiles that start from the 
same initial condition, that is, to different values of $\beta'(m_0)$.

If we are interested in the stationary path that touches the critical line tangent to it, we need to identify the values
of $\rho_I$ and the intersection point $m_I$ such that
\bse
\begin{equation}
    \frac{d}{d m} \beta (m_I, \rho_I) = 0
    \label{eq:onsager5}
\end{equation}
\begin{equation}
    \beta(m_I, \rho_I) = 0.
\end{equation}
\ese

By imposing the tangency condition (\ref{eq:onsager5}) in Eq. (\ref{eq:onsager2}), it is easy to find the relation between $\rho_I$ and $m_I$
\begin{equation}
    \rho_I = - \frac{\overline{L}_{ah}}{m_I \overline{L}_{aa}},
    \label{eq:onsager6}
\end{equation}
that we can use to replace $\rho$ in Eq. (\ref{eq:onsager4}), evaluated at $m = m_I$, to obtain 
\begin{equation}
\begin{split}
\beta(m_I, \rho_I(m_I))= \,& \beta_0 + \frac{\overline{L}_{aa}(\overline{L}_{ah}^2 (m_I-m_0) - 2 \overline{L}_{aa} \overline{L}_{hh} m_I)}{2 \overline{L}_{ah}^3} (m_I-m_0)  \\ &+\frac{\overline{L}_{aa}^2 \overline{L}_{hh}(\overline{L}_{ah}^2 - \overline{L}_{aa} \overline{L}_{hh}) m_I^2}{\overline{L}_{ah}^5} \ln\left(\frac{\overline{L}_{hh} \overline{L}_{aa} - \overline{L}_{ah}^2}{\overline{L}_{hh} \overline{L}_{aa} - \overline{L}_{ah}^2 \frac{m_0}{m_I}}\right).
\label{eq:onsager7}
\end{split}
\end{equation}

The value of $m_I$ can be determined by imposing the condition
$\beta(m_I, \rho_I(m_I))= 0$ on Eq.~(\ref{eq:onsager7}).

Once $m_I$ has been identified (by numerical methods), the parametric profile tangent to the critical
line in $m_I$ is the one given by Eq.~(\ref{eq:onsager4}), with $\rho = \rho_I$.

\subsection{The metastable path in the NT region}
\label{app.cubic}
From Eq. (\ref{eq:onsager4}) we can also obtain the metastable parametric profiles that enter in the localized phase, with $\beta<0$.
This time let us impose that the initial condition point $m_0$ is the point where $\beta'(m_0) = 0$, with $\beta_0 = \beta(m_0)$ which can be negative. Similarly to the previous subsection,
from Eq. (\ref{eq:onsager2}) we get $\rho = - \overline{L}_{ah}/(m_0 \overline{L}_{aa}) $, that we can replace in Eq. (\ref{eq:onsager4}) to obtain 
\begin{equation}
  \begin{split}
    \beta(m) = \,& \beta_0 - \frac{\overline{L}_{aa}(\overline{L}_{ah}^2 (m-m_0) + 2 \overline{L}_{aa} \overline{L}_{hh} m_0)}{2 \overline{L}_{ah}^3} (m-m_0)  \\ &+\frac{\overline{L}_{aa}^2 \overline{L}_{hh}(\overline{L}_{aa} \overline{L}_{hh} - \overline{L}_{ah}^2) m_0^2}{\overline{L}_{ah}^5} \ln\left(\frac{\overline{L}_{hh} \overline{L}_{aa} - \overline{L}_{ah}^2}{\overline{L}_{hh} \overline{L}_{aa} - \overline{L}_{ah}^2 \frac{m}{m_0}}\right).
  \end{split}
 \label{eq:onsager8}
\end{equation}
If we now expand the parametric profile $\beta(m)$ at $m_0$, we find
\begin{equation}
    \beta(m) =  \beta_0 + \frac{\overline L_{aa}\overline L_{ah}}{2 \det \overline L} (m - m_0)^2 + \frac{\overline L_{aa}^2\overline L_{ah} \overline L_{hh}}{3 (\det \overline L)^2 m_0} (m - m_0)^3 + {\cal O}(m - m_0)^4
\end{equation}

\subsection{Critical paths}
\label{app.cp}

Above the critical line, nonequilibrium paths give rise to the spontaneous emergence of peaks.
Here, we discuss a path lying along the critical line, under the additional assumption
of a non-constant energy flux: hence $j_h$ in Eq.~(\ref{eq:onsager}) is no longer constant.

Along the critical line $d\beta/dy = 0$ and the first equation in (\ref{eq:onsager}) becomes
\begin{equation}
\frac{dm}{dy} = -\frac{j_a}{\overline{L}_{aa}} m^2
\end{equation}
where both the flux $j_a$ and $\overline{L}_{aa}$ do not depend on $y$.
The general solution is (remember that along the $\beta=0$ line, $1/m = -a$)
\begin{equation}
\frac{1}{m}(y) =  \frac{j_a}{\overline{L}_{aa}} y + C
\end{equation}
where the integration constant $C$ and the value $j_a$ of the flux can
be determined by imposing the boundary conditions, $m(0)=m_0$, $m(1)=m_1$.
It is easily found that $C=1/m_0$ and
\begin{equation}
\label{eq.ja}
j_a = \left( \frac{1}{m_1}-\frac{1}{m_0} \right)\overline{L}_{aa}
\end{equation}
so that
\begin{equation}
\frac{1}{m(y)}  = \left ( \frac{1}{m_1}-\frac{1}{m_0}\right) y + \frac{1}{m_0}
\end{equation}
i.e. the mass-density profile is perfectly linear.

Now, knowing $m(y)$, we can focus on the equation ruling the energy flux $j_h$, which must be position
dependent to allow the path lie along the $\beta=0$ line.

\begin{equation}
j_h(y) = \frac{\overline{L}_{ah}}{m^3} \frac{dm}{dy} =
-\overline{L}_{ah} \left (\frac{1}{m_1}-\frac{1}{m_0} \right )  \left [ \left ( \frac{1}{m_1} - \frac{1}{m_0}\right) y
  + \frac{1}{m_0} \right ]
\end{equation}
Hence,
\begin{equation}
\label{eq.jhL}
j_h(0) = -\frac{\overline{L}_{ah}}{m_0} \left( \frac{1}{m_1}-\frac{1}{m_0} \right )
\end{equation}
\begin{equation}
\label{eq.jhR}
j_h(1) = -\frac{\overline{L}_{ah}}{m_1} \left( \frac{1}{m_1}-\frac{1}{m_0} \right )
\end{equation}
so that
\begin{equation}
\label{eq.Djh2}
\Delta j_h \equiv j_h(0) - j_h(1) =
\overline{L}_{ah} \left( \frac{1}{m_1}-\frac{1}{m_0} \right )^2 =
\overline{L}_{ah} \left( a_0 - a_1\right )^2
\end{equation}
as $a=-1/m$ (and $\overline{L}_{ah}\simeq 1.58$ as from 
Ref.~\cite{iubini23}).
Notice that the $N$ dependence here is hidden in implicit reference to a length 1. 

Finally, the ratio between the outgoing and incoming energy flux is
\begin{equation}
\frac{j_h(1)}{j_h(0)} = \frac{m_0}{m_1}  =\frac{a_1}{a_0}
\end{equation}

\section*{References}
\bibliographystyle{iopart-num}

\providecommand{\newblock}{}

\end{document}